\documentclass[manuscript,prologue,table,xcdraw]{acmart}
\usepackage{xcolor}
\usepackage[normalem]{ulem}
\usepackage{subcaption}

\usepackage{bm} 
\usepackage{float}
\usepackage{algorithm2e} 
\RestyleAlgo{ruled}
\SetKwComment{Comment}{/* }{ */}
\SetKwInOut{Require}{Require}
\SetKwInOut{Input}{Input}

\definecolor{highlightcolor}{rgb}{1, 1, 0.6} 

\newcommand{\st}[1] {}

\usepackage{listings}
\definecolor{codegreen}{rgb}{0,0.6,0}
\definecolor{codegray}{rgb}{0.5,0.5,0.5}
\definecolor{codepurple}{rgb}{0.58,0,0.82}
\definecolor{backcolour}{rgb}{0.95,0.95,0.92}
\definecolor{backcolour_prompts}{rgb}{0.92,0.91,0.62}

\lstdefinestyle{mystyle}{
    backgroundcolor=\color{backcolour},   
    commentstyle=\color{codegreen},
    keywordstyle=\color{magenta},
    numberstyle=\tiny\color{codegray},
    stringstyle=\color{codepurple},
    basicstyle=\ttfamily\footnotesize,
    breakatwhitespace=false,         
    breaklines=true,                 
    captionpos=b,                    
    keepspaces=true,                 
    numbersep=5pt,                  
    showspaces=false,                
    showstringspaces=false,
    showtabs=false,                  
    tabsize=2
}

\lstdefinestyle{prompts}{
    backgroundcolor=\color{backcolour_prompts},   
    commentstyle=\color{codegreen},
    keywordstyle=\color{magenta},
    numberstyle=\tiny\color{codegray},
    stringstyle=\color{codepurple},
    basicstyle=\ttfamily\footnotesize,
    breakatwhitespace=false,         
    breaklines=true,                 
    captionpos=b,                    
    keepspaces=true,                 
    numbersep=5pt,                  
    showspaces=false,                
    showstringspaces=false,
    showtabs=false,                  
    tabsize=2
}

\UseRawInputEncoding

\AtBeginDocument{%
  }

\setcopyright{acmlicensed}
\copyrightyear{2024}
\acmYear{2024}
\acmDOI{XXXXXXX.XXXXXXX}

\acmISBN{978-1-4503-XXXX-X/18/06}

\begin{document}

\title{Social Conjuring: Multi-User Runtime Collaboration with AI in Building Virtual 3D Worlds}


\author{Amina Kobenova}
\authornote{Authors contributed equally to this research and were affiliated with Microsoft .}
\email{akobenov@ucsc.edu}
\affiliation{%
  \institution{UC Santa Cruz}
  \country{USA}
}

\author{Cyan DeVeaux}
\authornotemark[1]
\email{cyanjd@stanford.edu}
\affiliation{%
  \institution{Stanford University}
  \country{USA}
}

\author{Samyak Parajuli}
\authornotemark[1]
\email{samyak.parajuli@utexas.edu}
\affiliation{%
  \institution{UT Austin}
  \country{USA}
}

\author{Andrzej Banburski-Fahey}
\email{abanburski@microsoft.com}
\affiliation{%
  \institution{Microsoft}
  \country{USA}
}

\author{Judith Amores Fernandez}
\email{judithamores@microsoft.com}
\affiliation{%
  \institution{Microsoft}
  \country{USA}
}

\author{Jaron Lanier}
\email{jalani@microsoft.com}
\affiliation{%
  \institution{Microsoft}
  \country{USA}
}


\renewcommand{\shortauthors}{Kobenova, DeVeaux, and Parajuli et al.}

\begin{abstract}
  Generative artificial intelligence has shown promise in prompting virtual worlds into existence, yet little attention has been given to understanding how this process unfolds as social interaction. We present \textit{Social Conjurer}, a framework for AI-augmented dynamic 3D scene co-creation, where multiple users collaboratively build and modify virtual worlds in real-time. Through an expanded set of interactions -- including social and tool-based engagements -- and spatial reasoning, our framework facilitates the creation of rich, diverse virtual environments. Findings from a preliminary user study (N=12) provide insight into the user experience of this approach, how social contexts shape the prompting of spatial environments, and perspective on social applications of prompt-based 3D co-creation. In addition to highlighting the potential of AI-supported multi-user world creation and offering new pathways for AI-augmented creative processes in VR, this article presents a set of implications for designing human-centered interfaces that incorporate AI models into 3D content generation.
\end{abstract}


\begin{teaserfigure}
  \includegraphics[width=\textwidth]{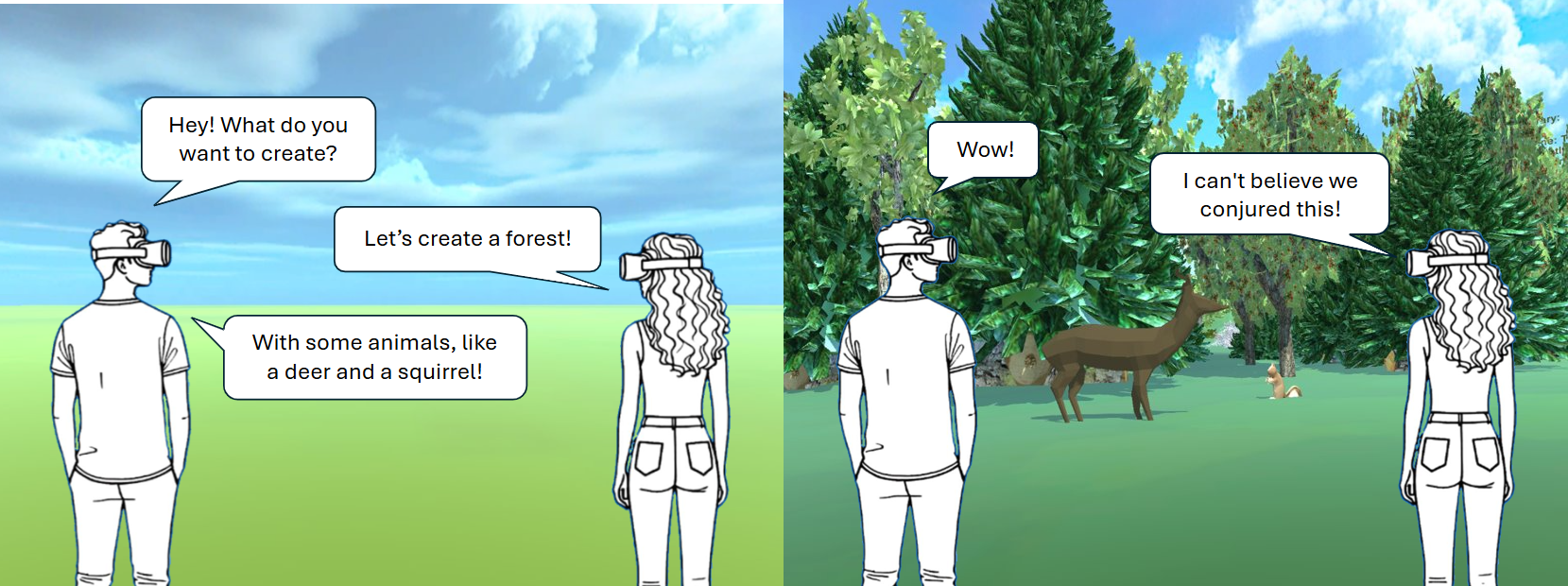}
  \caption{Collaborative Creation with \textit{Social Conjurer}: This image demonstrates a VR collaborative creation process where users generate a forest environment complete with animals, reflecting the system's capability for dynamic scene creation through prompts.}
  \Description{This image demonstrates a VR collaborative creation process where users generate an interactive forest environment complete with animals, reflecting the system's capability for dynamic scene creation through prompts.}
  \label{fig:teaser}
\end{teaserfigure}

\received{20 February 2007}
\received[revised]{12 March 2009}
\received[accepted]{5 June 2009}

\maketitle

\section{Introduction}

The prospects of turning virtual worlds into a collaborative communication medium stem from the early developments of virtual reality (VR) systems \cite{biocca1992insider, biocca1995virtual, schroeder1996possible}. Following a similar trend, modern immersive collaborative worldbuilding platforms, such as VRChat \cite{vrchat}, Minecraft \cite{minecraft}, Roblox \cite{roblox}, and The Second Life \cite{secondlife}, allow people to create shared environments and game simulations using pre-defined objects and tools in 3D.

Recent advancements in multi-user VR \cite{perlin2023multiperson, perlin2023sharedmr, doroudian2023collaboration} and artificial intelligence (AI) for VR \cite{de2024llmr, bozkir2024cui, giunchi2024dreamcodevr} have the potential to significantly influence the way we interact and collaborate in virtual 3D worlds. Despite these developments, the integration of generative AI (GenAI) tools and agents into collaborative VR environments remains understudied. GenAI driven tools can enhance collaborative experiences when creating 3D scenes, objects, and scripts in real-time. However, collaborative runtime VR environments, especially those utilizing Large Language Models (LLMs) and Vision Language Models (VLMs), face technical networking limitations and are still in the early stages of development, e.g. \cite{ai-networking}.

The aim of this research is to present a framework for the Human-AI co-creation of multi-user, collaborative and modifiable virtual worlds at runtime. While previous work has demonstrated the feasibility of generating virtual scenes leveraging text-based prompts, our work expands this by augmenting prompt-based interaction, moving from the creation of static scenes to more dynamic environments, and considers the experience of multiple users immersed within the same VR environment. We accomplish this by broadening the number of interactions afforded to the user (e.g., social interaction, tool-based interaction), while also improving the diversity and quality of the scenes produced. Overall, our research was driven by the following questions:

\begin{itemize}
    \item \textbf{RQ1:} How can we develop a real-time, collaborative, spatially-aware system, integrating both large language models (LLMs) and vision language models (VLMs), to facilitate the co-creation of virtual worlds?
    \item \textbf{RQ2:} How do shared virtual environments influence spontaneous, collaborative world building?
    \item \textbf{RQ3:} What are the challenges and opportunities of using language-based prompts to generate 3D virtual environments?
\end{itemize}

In this paper, we present several key contributions to the field of Human-Computer Interaction (HCI). First, we introduce \textit{Social Conjurer}, a novel system that integrates multiplayer functionality with AI-driven world building, enabling spontaneous and collaborative creation of 3D content. Second, we present study findings on how social, multi-user contexts shape prompting spatial scenes into existence. Our study highlights the diverse motivations and patterns of interaction among users, examines how discrepancies between user intent and AI interpretation shape scene creation, and considers user perspectives on the future social applications of prompt-based 3D creation. Additionally, we contribute to the field by discussing design implications and proposing future research directions for similar systems, emphasizing the importance of accessibility, ethical considerations, and optimizing user experience in collaborative virtual environments.

\section{Related Work}

To contextualize our work, we first describe recent progress in using AI for generating 3D content. Next, we review prior work on spatial reasoning for virtual layout design. Finally, we ground the multi-user focus of this paper in a review of collaborative world building in VR.

\subsection{Creating 3D Content with AI} 
The recent growth in popularity of GenAI can be in part attributed to ongoing advancements in the Human-AI interaction space \cite{shi2023hci}. The development of interfaces that enable users to engage with GenAI (via prompts \cite{openai2024chatgpt, liu2022design} and other input modalities \cite{lin2020your}) has contributed to a rise in Human-AI created content, including within 3D domains \cite{de2024llmr, yang2024holodeck}. For example, past scholars have used GenAI in the pipeline of generating 3D assets \cite{richardson2023texture, liu20233dall, faruqi2023style2fab, tochilkin2024triposr}. Relevant to this work, popular AI models, such as LLMs and VLMs, have been integrated into user interfaces that facilitate the spatial embodiment of AI through the creation and modification of 3D content \cite{shi2023hci, hong20233d, roberts2022stepspromptbasedcreationvirtual, huang2024realtimeanimationgenerationcontrol}. This progress has helped reduce barriers in authoring 3D content, which can be challenging due to the required expertise in using 3D modeling software and the time-consuming nature of it \cite{nebeling2018trouble, ashtari2020creating}.

Building on past work exploring language-driven generation of 3D scenes  \cite{ma2018language, seversky2006real, coyne2001wordseye, chang2017sceneseer}, a recent research agenda in the field of HCI and AI has emerged in understanding how LLMs can be used to aid this process of producing virtual environments. De La Torre, Fang, Huang, and colleagues \cite{de2024llmr} presented \textit{LLMR}, a system that enabled the real-time generation and modification of interactive scenes in Mixed Reality (MR) using language prompts. Yang and colleagues \cite{yang2024holodeck} developed \textit{HOLODECK}, a system that similarly used prompts as input for fully automated 3D environment creation. Öcal and colleagues \cite{ocal2024sceneteller} presented \textit{SceneTeller}, which took user descriptions of object placement as input and generated corresponding 3D scenes. Manesh et al. \cite{aghel2024people} \textit{Ostaad} system is a prototype for a conversational agent that allowed in-situ design of interactive VR scenes. While this prior work has demonstrated the feasibility of leveraging LLMs for Human-AI collaboration in generating of virtual scenes, there is a lack of work exploring how these systems facilitate collaboration between multiple humans and AI in pursuit of the same goal. Our current system and evaluation aims to fill this gap by presenting a multi-user framework for language-driven 3D scene creation.

Parallel to this work has been a small number of scholars exploring the user experience of prompting of 3D scenes and how it shapes the spatial design process. In an elicitation study examining users' expectations when prompting interactive VR scenes, Manesh and colleagues' \cite{aghel2024people} found that participants expected for the AI-driven tool to have embodied knowledge of the spatial arrangement of the environment, an understanding of objects in the scene, and memory of prior prompts. Lee and colleagues \cite{lee2024sketch} found that within design work flows, prompting 3D objects promoted divergent thinking whereas sketching 3D objects promoted convergent thinking. We extend this work by using our system as a design probe for gathering user perspectives on the future utility of collaborative prompt-based scene creation.

\subsection{Spatial Reasoning and Scene Layouts}
Spatial reasoning involves the ability to comprehend and interpret the relationships between objects and spaces in a given environment. It requires understanding how objects are positioned, oriented, and arranged relative to one another, as well as how they interact within a space. AI systems that incorporate spatial reasoning are often used in robotics \cite{Gubbi2020SpatialRF}, gaming \cite{Bergsma2008AdaptiveSR}, and augmented/virtual reality applications \cite{Papakostas2024}. 

 Exploration of spatial reasoning in large transformer architectures was conducted by \cite{Bang2023AMM} in which they evalute ChatGPT on two spatial reasoning tasks: SpartQA \cite{Mirzaee2021SPARTQAAT} and StepGame \cite{Shi2022StepGameAN}. These datasets consist of story-question pairs written in natural language about $k$ relations
of $k+1$ ($k \leq 10$) entities, finding that off the shelf approaches like ChatGPT did not fare well on this task.

Automated layouts of scenes composed out of multiple 3D assets have been recently explored in several settings. SceneCraft \cite{hu2024scenecraft} introduced an VLM agent capable of synthesizing Blender code for iterative scene composition, in which the system learned to implement spatial constraints over multiple iterations of proposed layouts with visual feedback. Separate streams of work have considered making the layouts learnable. In \cite{epstein2024disentangled} the authors have jointly trained 3D Neural Radiance Field (NeRF) models for each asset and then exposed their positions and rotations as differentiable parameters to be jointly optimized via gradient descent. Generating full scenes with NeRFs has also been considered in \cite{cohen2023set, po2024compositional}. The authors of \cite{zhou2024gala3d} first used an LLM to generate an initial layout and then used a layout-guided 3D Gaussian representation with adaptive geometric constraints. Gaussian splatting methods for scene generation have also been considered in prior work \cite{yuan2024dreamscape, li2024dreamscene, ocal2024sceneteller, zhang20243ditscene, chung2023luciddreamer}.

\subsection{Collaborative Worldbuilding in 3D and VR}

Multi-user capabilities form a critical factor in shaping gaming and immersive 3D experiences. Collaborative play, worldbuilding, and their implications have been explored through 3D gaming platforms like Minecraft and Roblox in social and educational contexts \cite{hughes2024minecrafters, volum-etal-2022-craft, davies2024roblox, Wardhana_2021, queiroz2023collaborative}, in AI-supported world generation \cite{35, earle2024dream, tim2023minecraft}, and in fostering creativity and problem-solving skills \cite{jawhar2024collaborative, fan2022open, hjorth2021exploring}.  These platforms provide opportunities for both guided and open-ended collaborative creation, allowing users to engage in synchronous and asynchronous tasks that encourage co-creation in dynamic environments. While many of these works examine technical and collaborative approaches to 3D worldbuilding, little literature emphasizes runtime collaboration supported by AI-agents in shared immersive environments. This gap highlights the need for further investigation into how AI can enhance real-time multi-user collaboration in 3D worlds, particularly by adapting to user input in complex, evolving virtual spaces.

While 3D world co-creation has been dominating gaming platforms, shared virtual reality spaces, specifically referred to as ``Social VR'' \cite{li2020social, osborne2023social}, have been exploring collaborative task performance, presence, and virtual embodiment. Social VR environments extend beyond traditional gaming, offering immersive opportunities for interpersonal interaction and co-presence, where users can engage with one another and virtual content simultaneously. Applications of social VR environments have been prominent in educational contexts \cite{drey2022vr, dunmoye2024exploratory, han2023people, deveaux2022learning}, body and avatar representation \cite{freeman2021social}, and design implications for collaborative VR \cite{kim2024behavioral}. However, similar to multiplayer gaming experiences, research exploring collaborative, multi-user world creation in VR at runtime has been limited. This is a crucial aspect for further study, as it presents opportunities to investigate the integration of co-presence, real-time editing, and immersive interaction in a way that transcends traditional, turn-based collaborative models.

Moreover, research on collaborative worldbuilding in VR has predominantly focused on static, pre-designed environments where interactions are limited to object manipulation or user-driven scenario creation \cite{wang2021authenticity, plass2022designing, zaker2018virtual}. While these systems have proven effective in educational and social settings, the potential of real-time, dynamic worldbuilding remains underexplored. Real-time co-creation, supported by AI, could enable users to modify both the environment and its underlying rules as interactions unfold, opening new dimensions for collaborative work in virtual spaces. By introducing AI-driven features into the process of worldbuilding, e.g. \cite{de2024llmr}, researchers can explore how AI agents assist in decision-making, environmental scaling, and adaptive scene generation based on the behavior and goals of participants. This would not only enrich user interaction but also enable complex, evolving environments that adapt to the collaborative efforts of multiple users simultaneously. Expanding upon these ideas, our research aims to explore new paradigms in multi-user runtime worldbuilding, integrating the human-AI interaction in 3D creation process.

\section{Design Goals}
Motivated by the social affordances of VR, potential of LLMs for scene generation, and limitations of prior work, we sought to address existing gaps in co-creative embodied AI experiences at runtime. We built \textit{Social Conjurer}, an interface that enables the real-time, collaborative creation of virtual worlds using language-driven prompts and sketches. In doing so, we identified three key design goals to incorporate into our system: (1) AI-Augmented Creation of Virtual Scenes at Run-Time, (2) Embodied Interaction with Scenes in VR, and (3) Shared Virtual Spaces for Collaborative Worldbuilding.

\subsection{AI-Augmented Creation of Virtual Scenes at Run-Time}
A core requirement of our system was to enable the creation of dynamic, virtual scenes at run-time. By leveraging LLMs and VLMs, we sought to provide users with a flexible experience that would allow them to prompt for any content, behavior, or interaction they envisioned within their scene. We aimed for sketches and language to be used as a primary methods for users to prompts scenes into existence. Acknowledging the cognitive load associated with crafting language-based prompts \cite{dang2022prompt, zamfirescu2023}, we aimed to minimize the need for users to provide elaborate instructions to our system. Our system should be able to infer details about a scene's environment, objects, and their placement based on a user-provided prompt, regardless of its level of detail. Therefore, we sought to design a system with spatial reasoning and embodied knowledge. At the same time, there is value to iterative refinement in language-guided 3D scene generation \cite{aghel2024people}. Hence, our intended approach would also allow users to modify their scene with additional prompts, rather than relying on a fully AI automated pipeline \cite{yang2024holodeck}. 

\subsection{Embodied Interaction with Scenes in VR}
To enhance the immersive potential of our system, another design goal was to enable embodied interaction with the user-generated scenes in VR. While previous systems that have allowed users to generate virtual scenes with language prompts have overlooked consideration for users' immersion in the generated scenes \cite{yang2024holodeck, de2024llmr}, VR offers unique opportunities for embodied interaction. Gesture and body-based interactions, facilitated by body tracking, are among the most compelling affordances of the medium \cite{johnson2018immersive} and should be considered within our framework. Therefore, we aimed to address this by giving users the ability to navigate and interact with their generated scene through desktop and in VR. Moreover, we sought to allow users to prompt interactive elements into their scene, using VR controllers as input. In doing so, we aimed to enable creation of scenes that are not only visually appealing, but also engaging to be in.   

\subsection{Virtual Spaces for Collaborative Worldbuilding}

In line with the multi-user vision of our \textit{Social Conjurer} system, another key design goal was to create shared virtual spaces for collaborative world building. Runtime collaboration built through game engines, in our case Unity 3D \cite{unity}, requires a number of computational resources and technical capabilities. These include real-time synchronization of multiple users, efficient network handling, and the ability to dynamically instantiate and modify objects -- known as \textit{prefab assets} -- in a shared environment. Additionally, ensuring low-latency communication and seamless integration of user inputs across distributed systems demands advanced processing power and optimized algorithms for managing concurrency in multi-user settings \cite{engelbrecht2021building}. 

Multiple networking frameworks and packages exist to support the multiplayer functionality: Unity Multiplayer Networking \cite{unity-multiplayer-networking},  Photon libraries \cite{photon-fusion, photon-engine}, and open-source research releases, such as the ``Metaverse Unity Networking'' \cite{ai-networking}. However, these networking libraries are restricted by instantiating pre-registered 3D assets in Unity scenes. Any instantiated assets are required to be pre-defined in the Unity Editor and must be referenced in the respective networking library, such that the next time the program is running, the library can clone those objects for other clients in the network. In real-time collaborative environments, however, registering 3D objects is infeasible, as they do not exist inside the Unity Editor ahead of time.

To overcome the limitations of pre-registered 3D assets and enable true real-time worldbuilding, a more flexible and dynamic system is required. Our proposed solution addresses these challenges by integrating AI-supported generation and runtime instantiation of 3D assets directly into the Unity environment, allowing users to collaboratively create and modify virtual worlds without the constraints of pre-defined objects. In the following section, we detail the architecture and technical components of this approach, which facilitates seamless, multi-user co-creation in shared virtual spaces.

\section{System Description}

\begin{figure}[ht]
  \centering
  \begin{subfigure}[t]{.32\linewidth}
    \centering
    \includegraphics[width=\linewidth]{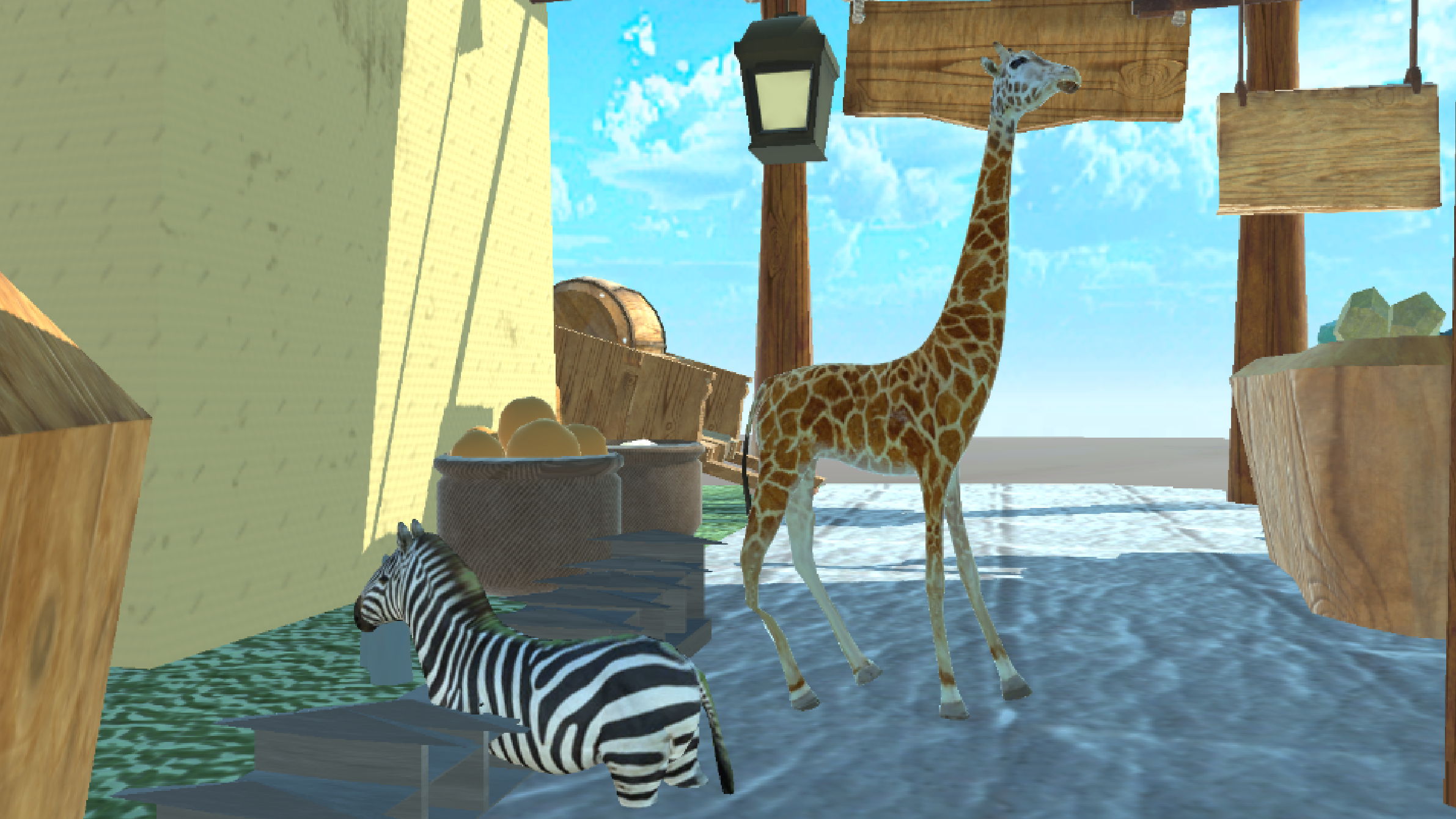}
    \caption{Scene generated by the prompt \textit{"create a zoo"}.}
  \end{subfigure}
  \hfill
  \begin{subfigure}[t]{.32\linewidth}
    \centering
    \includegraphics[width=\linewidth]{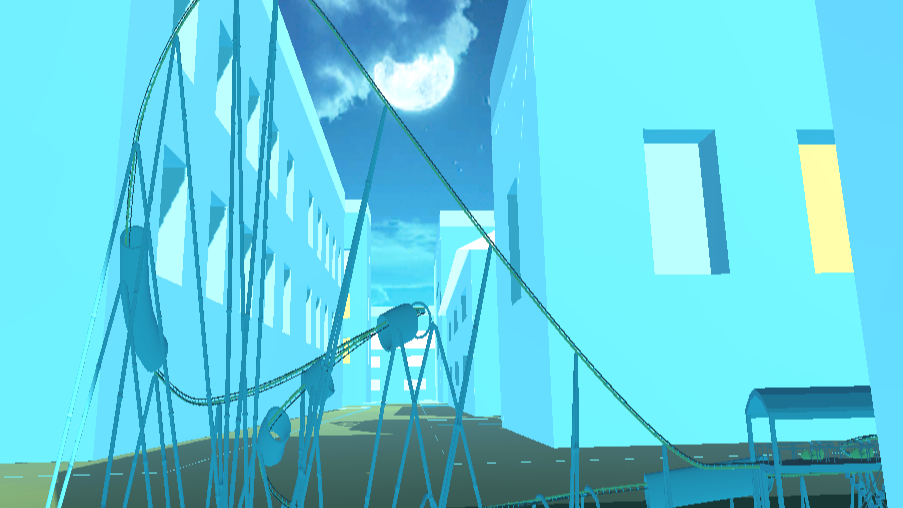}
    \caption{Scene generated by the prompt \textit{"create an amusement park at night"}.}
  \end{subfigure}
  \hfill
  \begin{subfigure}[t]{.32\linewidth}
    \centering
    \includegraphics[width=\linewidth]{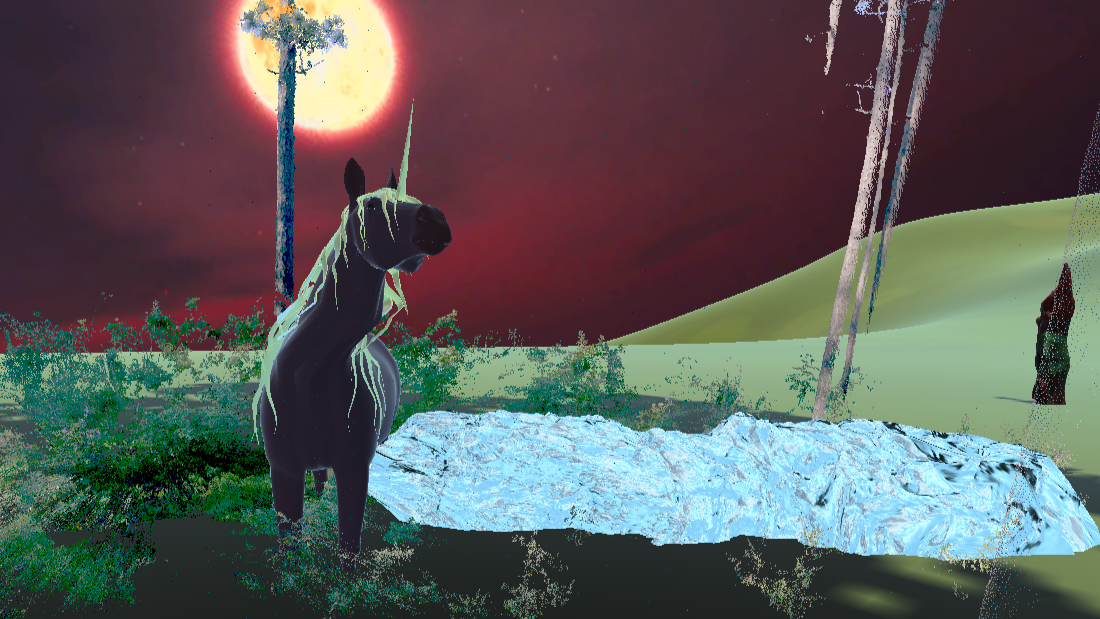}
    \caption{Scene generated by the prompts \textit{"create a fantasy forest"} and \textit{"add a unicorn and gnome,"} and modified with tools.}
  \end{subfigure}
  
  \caption{Examples of scenes generated by Conjurer.}
  \label{fig:scene-examples}
  \vspace{-6mm}
\end{figure}

We designed the \textit{Social Conjurer} system by building on top of the open-sourced LLMR architecture proposed in \cite{de2024llmr}, which is effectively a chain of LLM calls with feedback loops between two of the agents (namely Builder and Inspector). In the LLMR system a user prompt is first passed to a Scene Analyzer agent, which extracts relevant virtual scene information and passes it to a code writing Builder agent (equipped with a library of skills), whose output is inspected by an Inspector agent (and sent back to the Builder for improvements in a feedback loop), before being passed to a runtime Roslyn C\# compiler, resulting in a creation or an update of a virtual scene (See Figure ~\ref{fig:scene-examples}). In this work, we expand on this architecture in several ways, with the goals of improving on the quality of generated worlds, adding networked multiplayer capabilities and making the system accessible to non-expert users. 

\begin{figure}[ht!]
    \centering
    \includegraphics[width=1\linewidth]{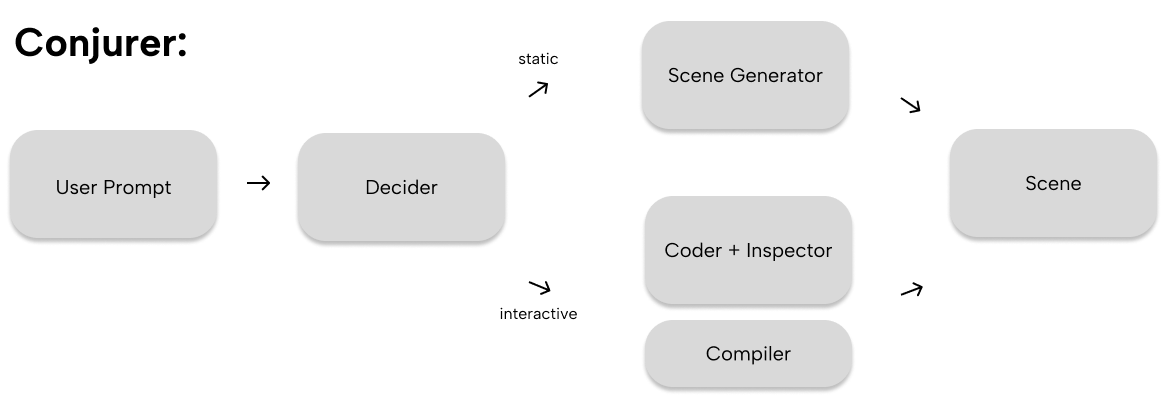}
    \caption{Overview of the single-player Conjurer architecture}
    \label{fig:conjurer}
\end{figure}  

\vspace{1cm}
At a high level, \textit{Social Conjurer} takes a prompt from one of the users and creates or changes a shared networked virtual world. For clarity of presentation, we can first introduce the concept of a single-user \textit{Conjurer} system (Fig. \ref{fig:conjurer}), and then discuss the adaptation of it to the multi-user setting. Here, the user prompt is first passed to Decider -- an LLM agent that decides whether satisfying this request requires writing custom code to implement dynamic interactive elements, or whether it can be satisfied with simple LLM-powered function calls to create static environments. This is a crucial distinction that allows for a much more rapid scene generation. Always requiring writing of code to implement the visual scene and then inspecting it in a feedback loop (as is done in LLMR), leads to very long wait times before a successful code candidate is found, and often is visually subpar. In the case where the request does ask for a static scene without any interactive elements (at least initially), we pass the request to the Scene Generator module, which we describe below. This module acts asynchronously in generating the overall environment, finding relevant assets and then placing them in the scene over a few iterations, presenting the intermediate results to the user as a growing virtual world that improves over time. If the user prompt does ask for interactive scenes, we pass it to a modified version of LLMR, one in which the Scene Generator API is exposed to the coding agent, together with improvements on tool generations discussed in Section \ref{sec:tools}. This still involves a Scene Analyzer, Coder (which we rename from Builder, to clarify that it is meant for coding interactive elements now, rather than just building scenes) and an Inspector.

\begin{figure}
    \centering
    \includegraphics[width=1\linewidth]{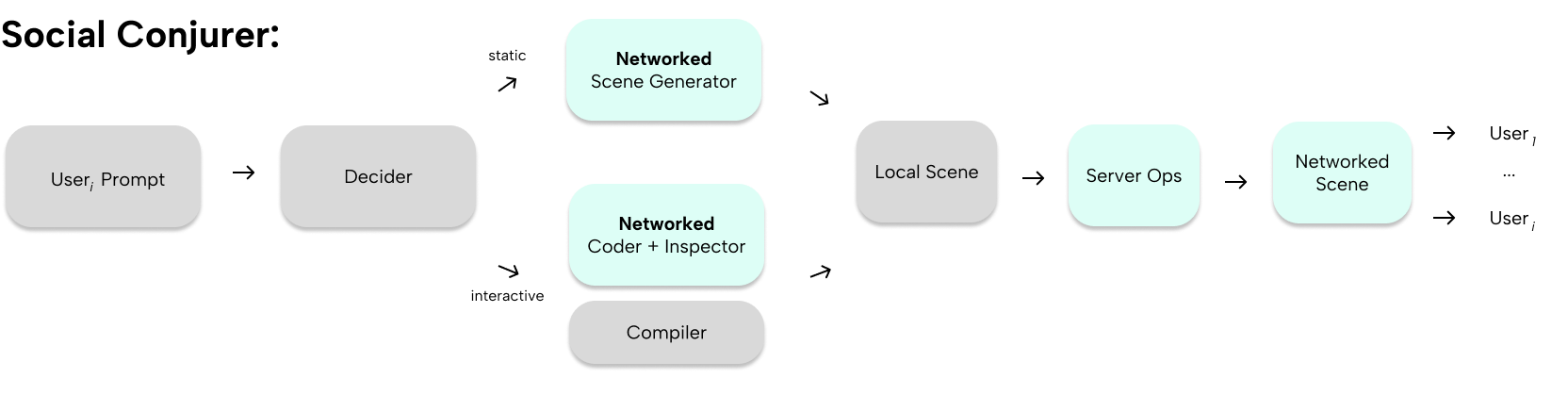}
    \caption{High level overview of Social Conjurer. Several of the submodules are themselves multi-agent architectures, described in following subsections.}
    \label{fig:social-conjurer}
\end{figure}

We are now ready to introduce the \textit{Social Conjurer} as a generalization of the \textit{Conjurer} architecture above, see Fig. \ref{fig:social-conjurer}. First, we consider the setting in which we have $n$ users: $\textnormal{user}_1,\ldots,\textnormal{user}_n$. We then pass a prompt from some $\textnormal{user}_i$ to a Decider that routes the request to the same modules as above, with the difference that their outputs are Networked Objects in the Unity engine, with Remote Procedure Calls (RPCs) announcing their properties. This results in a scene local to the $\textnormal{user}_i$, which is then passed to the Networking module. This module registers and bakes the various objects and tools, thus creating a shared networked scene that is communicated to all the other users. For details of this networking logic, see Section \ref{sec:networking}.

\subsection{Scene Generator: Spatial Reasoning and Environment Generation}
The Scene Generator module is used when a user prompts for a scene that can be built out of static objects. This module is composed out of two submodules, one responsible for generating the environments and terrains, the other responsible for obtaining 3D assets and reasoning on their placement and sizing within the scene. Each of these two submodules are called in parallel, and are themselves asynchronous multi-agent architectures. The \textit{Environment Generation} submodule decides on the type of terrain to use, how to texture it, whether to add a Skybox or a body of water to it, and is described in detail in Section \ref{sec:environment}. The \textit{Spatial Reasoning} submodule asynchronously proposes relevant assets to be added to the scene and retrieves them (Section \ref{sec:asset-retrieval}), while also deciding on their initial layout in the scene (Section \ref{sec:initial-layout}). Because these two processes are done in parallel, their results are presented to the user independently, which reduces the perceived latency of the system. After these two processes conclude, we can iterate on an improved layout and orientations of the assets in the scene, described in Section \ref{sec:layout-improvement}. To iterate on the layout, the submodule could either work in place and use the Unity renderer to gather visual feedback, or use an external lightweight rendering engine, similarly to a sketchpad \cite{hu2024visual}. Here, we decided on the latter to reduce the amount of processing on the clients, and we run a Python flask server with Matplotlib \cite{Hunter:2007} and Trimesh \cite{trimesh}, which communicates results to the Unity engine.


\subsubsection{Environment Generation}
\label{sec:environment}

\begin{figure}
    \centering
    \includegraphics[width=1\linewidth]{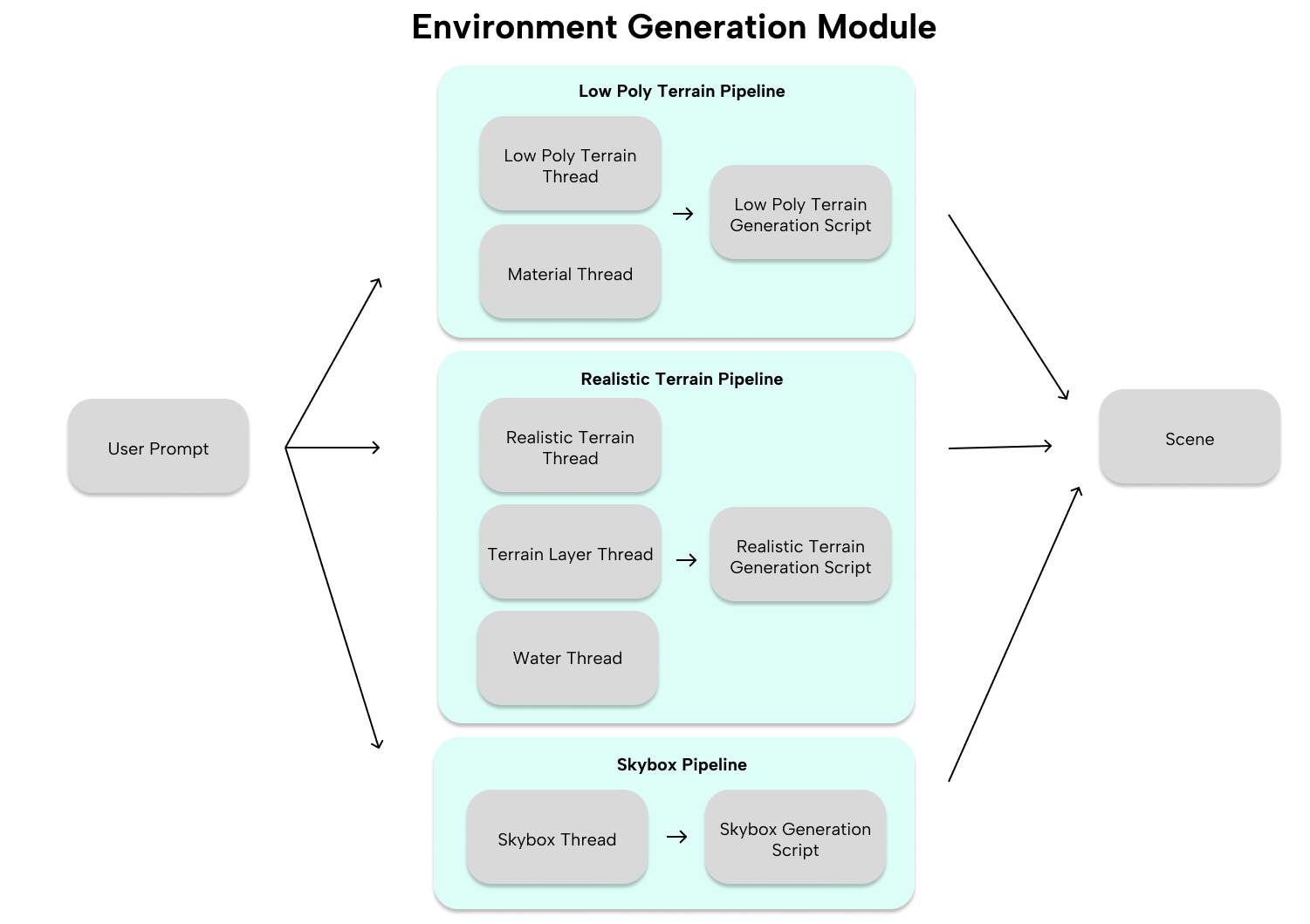} \pdfstringdef{\alttext}{This diagram illustrates the environment generation pipelines for a virtual reality system. There are three main pipelines: Low Poly Terrain Pipeline, Realistic Terrain Pipeline, and Skybox Pipeline. Each pipeline starts from a user prompt and results in generating a scene. In the Low Poly Terrain Pipeline, two threads—'Low Poly Terrain Thread' and 'Material Thread'—feed into the 'Low Poly Terrain Generation Script'. The Realistic Terrain Pipeline consists of three threads—'Realistic Terrain Thread', 'Terrain Layer Thread', and 'Water Thread'—that connect to the 'Realistic Terrain Generation Script'. Finally, the Skybox Pipeline has a single 'Skybox Thread' that connects to the 'Skybox Generation Script'. All pipelines ultimately lead to the scene creation process.}
    \caption{Environment Generation Module for the \textit{Social Conjurer} System.}
    \label{fig:envr-gen-diagram}
\end{figure}
Our environment generation submodule was designed to dynamically generate virtual environments based on user's prompts (See Figure ~\ref{fig:envr-gen-diagram}). This module uses the OpenAI Threads API to asynchronously generate separate elements of the environment  in parallel, including the terrain, terrain surface properties, skybox, and water. Separate agents are created for handling each environmental component, based on the users prompts. If the system is unable to complete the task, it retries up to 3 times. Once responses from each thread are received, the outputs of each agent are used as inputs for pre-written environment generation methods that load the separate components into the Unity scene. 

To handle terrain-generation related tasks, we created two different potential pipelines depending on the desired level of realism. A \textit{Low Poly Terrain Selection} agent uses the prompt to determine the most suitable terrain type from a predetermined set of options (e.g, \textit{"farmland," "mountain"}). Each terrain type corresponds to a set of noise parameters that are used to modify the surface of a terrain loaded into the scene based on the agent's output. A \textit{Realistic Terrain Selection} agent on the other hand analyzes the prompt to select a corresponding real-world location with a terrain suitable for the user's description. The proposed coordinates are used as input for a script that uses the Mapbox API \cite{mapbox} to obtain real-world elevation that is then converted into a height map and used to generate a terrain in the scene.  A \textit{Material Selection} agent selects the most appropriate option, given a user's prompt, from a predefined list of material names (e.g., \textit{"grass," "sand"}). As each material in the list exists within the project's resources, the specific material is applied to the low poly terrain with a pre-written method. Similarly, the \textit{Terrain Layer Selection} assistant selects the most appropriate terrain layer texture from a pre-defined list of names (e.g., \textit{"Grass\_TerrainLayer," "Sand\_TerrainLayer"}). This output is used as input for applying an existing terrain layer texture to the realistic terrain. The \textit{Skybox Selection} agent responds with the most appropriate skybox from a list of skybox asset names (e.g., \textit{"daytime\_bright\_skybox," "sunrise\_cool\_skybox"}). This information is used as input to a skybox loader script. Finally, the \textit{Water Selection} agent decides whether or not it would be appropriate to add a water prefab into the scene, and runs the water loading method accordingly.

\begin{figure}[ht]
  \centering
  \begin{subfigure}[b]{.48\linewidth}
    \centering
    \includegraphics[width=\linewidth]{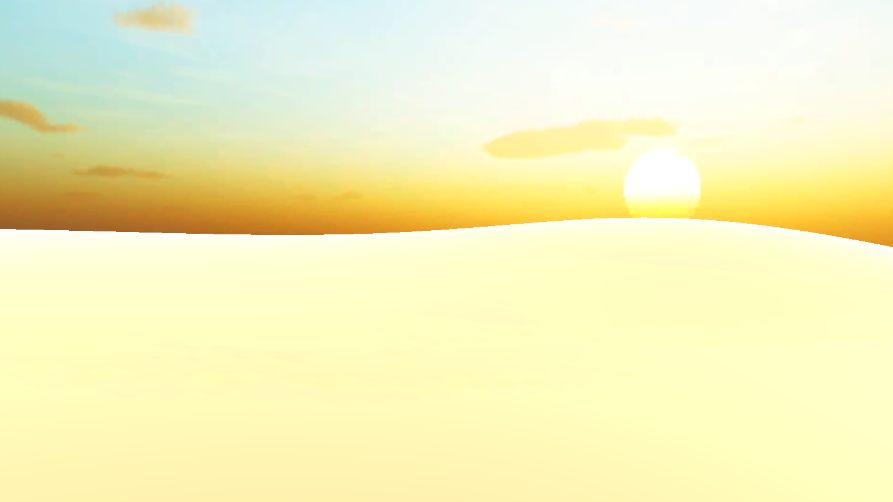}
    \caption{Low poly environment generated by the prompt \textit{"create a desert scene on a cool morning"}.}
  \end{subfigure}
  \hfill
  \begin{subfigure}[b]{.48\linewidth}
    \centering
    \includegraphics[width=\linewidth]{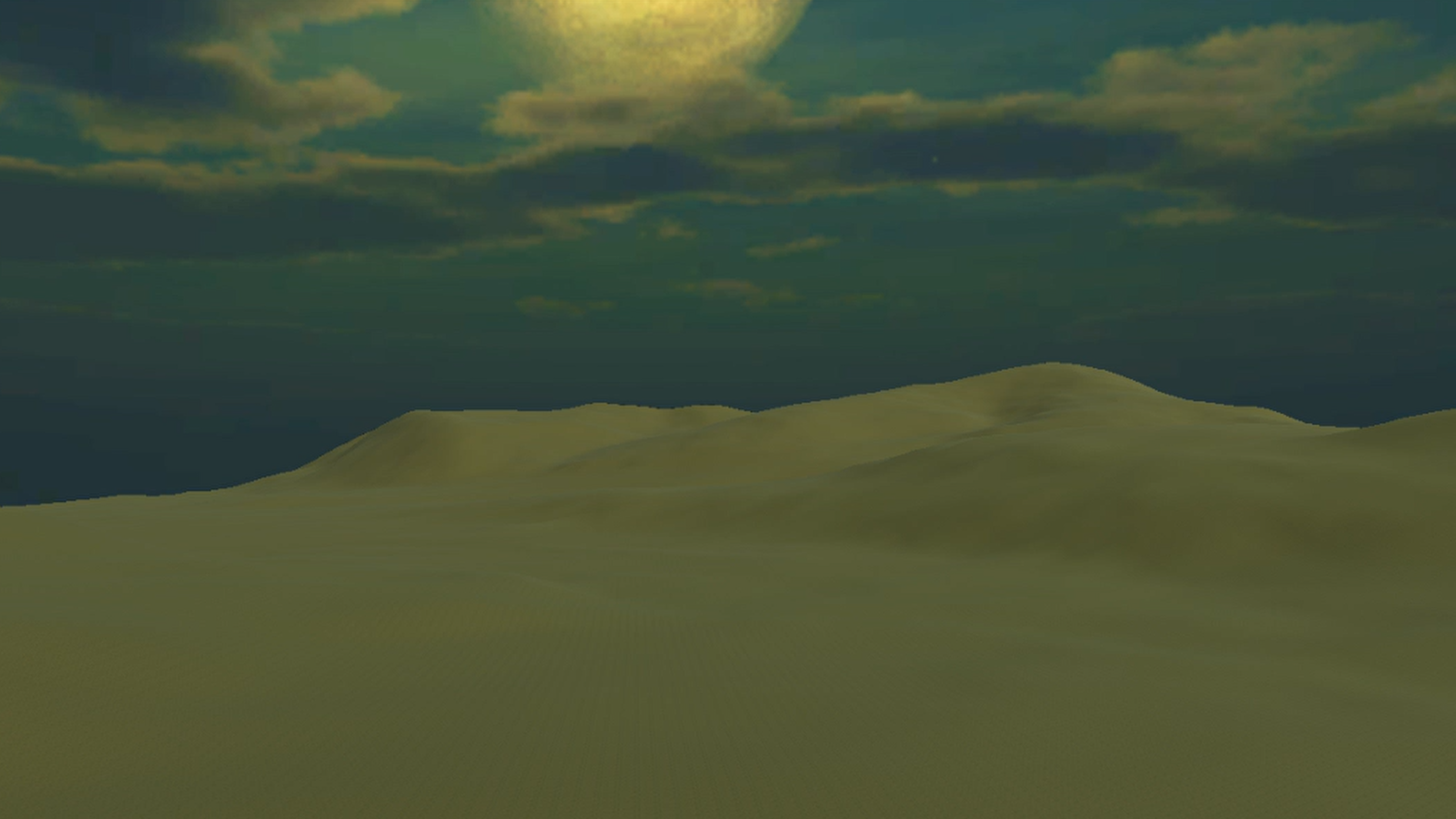}
    \caption{Realistic environment generated by the prompt \textit{"create a grassy field at night with gentle hills"}.}
  \end{subfigure}
  \caption{Examples of environments created with our environment generation module.}
  \label{fig:terrain-examples}
  \vspace{-6mm}
\end{figure}

By completing each of these tasks, a full environment is loaded into the scene (See Figure ~\ref{fig:terrain-examples}). Although this module can be called for any prompt, our application of it was based on the first prompt a user would input into the system. Prior to this, the environment consists of a default scene.

\subsubsection{Asset Proposal and Retrieval}
\label{sec:asset-retrieval}

The first part of this pipeline involves determining suitable assets for the given scene. Users have the freedom to directly ask for specific assets, or give a high level overview and ask for a particular number of assets they want in the scene, or leave both up to the system's discretion. We prompt an LLM (in our experiments we used GPT-4o \cite{gpt4o}) with examples of scenes and corresponding assets and parse the response to determine which assets to use. This first step results in the names of the assets, which we then display to the user as red 1x1x1 wireframe boxes with asset names floating inside placed around the scene without overlapping.

\begin{figure}[ht]
  \centering
  \subcaptionbox*{Precomputation stage}{%
    \includegraphics[width=.5\linewidth]{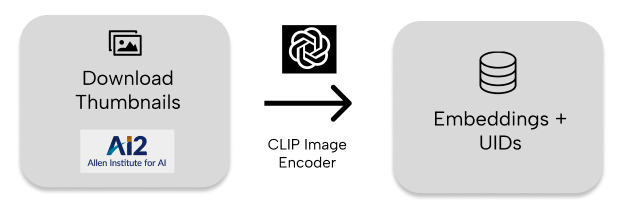}%
  }
  \vspace{2mm}
  \subcaptionbox*{Runtime stage}{%
    \includegraphics[width=1\linewidth]{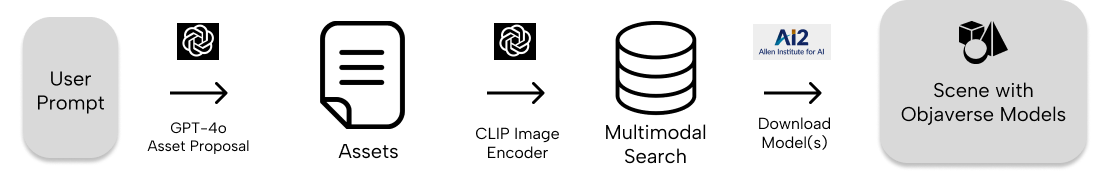}%
  }
  \caption{Asset proposal and retrieval logic.}
  \label{fig:retrieval-logic}
  \vspace{-6mm}
\end{figure}

To actually retrieve the meshes, we leverage Objaverse \cite{objaverse}, a collection of \textasciitilde{}800K annotated 3D objects. Objaverse has a `tag' key that provides coarse grained labels for particular objects. However, we found that these annotations are not always reliable. To improve the recall quality of meshes, as well speed up the asset retrieval process, we devised a method that leverages the CLIP \cite{Radford2021LearningTV} model for multimodal semantic search. We achieved this by pre-downloading thumbnails of all the assets and saving a compact CLIP embedding representation to be used for inference. Thus, when we query for a particular asset, we dynamically embed the text and search over pre-computed embeddings to return the unique identifiers (UIDs) for the particular meshes we want to download. Once each mesh gets downloaded, we place it inside the associated wireframe box, rescale it uniformly to fit within the 1x1x1 confines of it, and delete the now unnecessary label. This process is visualized in Figure \ref{fig:retrieval-logic}.

\subsubsection{Layout Arrangement}
\label{sec:initial-layout}

Once we have the relevant asset names, we can obtain their proposed layout in the scene in parallel to actually loading them into the scene. We prompt an LLM with a few in-context examples that take as input a scene description and asset names, and output proposed sizes of these objects (the x, y, z extents of their bounding boxes), their positions and orientations in the scene. We found that these in-context examples, combined with a set of guidelines on grounding the assets in an coordinate axis grid proved to be very effective in unleashing the base LLM's ability to reason about spatial arrangements, resulting in a reasonable first proposal at a scene layout. For details of the prompting strategy, see the Supplementary Material. 

Once we obtain the proposal layout from the LLM, we parse it and then apply the suggested changes to each asset. We appropriately rescale the wireframe bounding boxes, shift them to the proposed positions and then change the wireframe colors to yellow, to indicate that the first pass of laying out the scene has been completed, but more is still to come. Note, that once the appropriate assets are loaded into the containers, we make sure to rescale them uniformly to the largest suggested extent at this step, to avoid having the actual assets be unnaturally squished. We then store an updated dictionary of assets with their meshes, first pass of positions and rotations, and updated sizes.


\begin{figure}[ht]
  \centering
  \begin{minipage}[b]{0.7\linewidth}  
    \centering
    \subcaption{Final Simplified Rendering}
    \includegraphics[width=\linewidth]{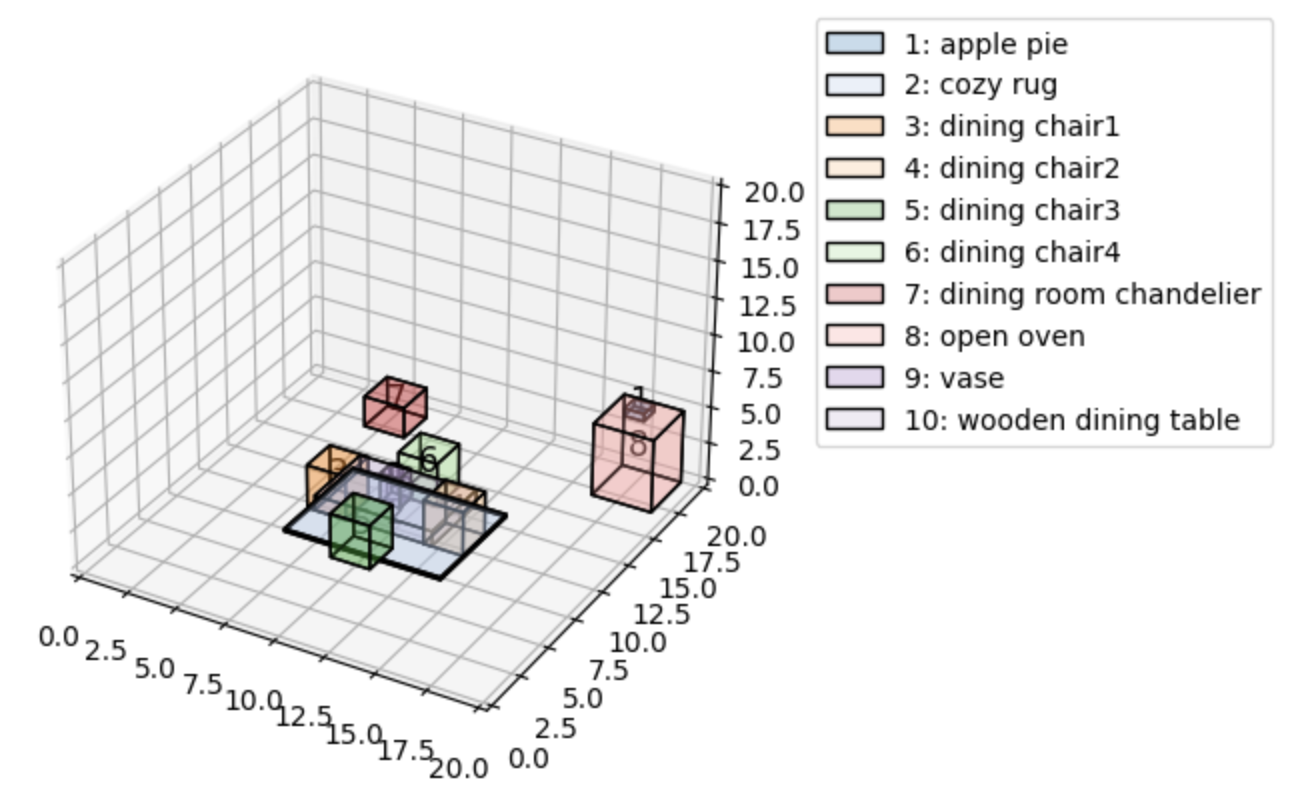}
  \end{minipage}%
  \hfill
  \begin{minipage}[b]{0.3\linewidth}  
    \centering
    \subcaption{Trimesh Wireframes}
    \includegraphics[width=\linewidth]{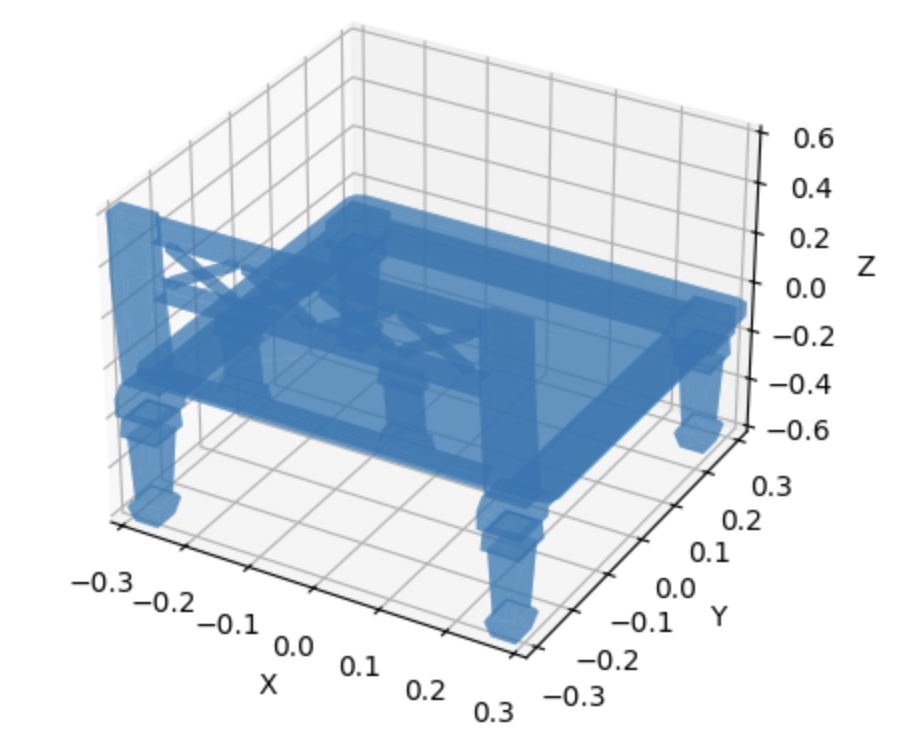}
  \end{minipage}
  
  \vspace{2mm}
  
  \begin{minipage}[b]{0.5\linewidth}  
    \centering
    \subcaption{Coordinate axis grounded visuals for semantic scene creation}
    \includegraphics[width=\linewidth]{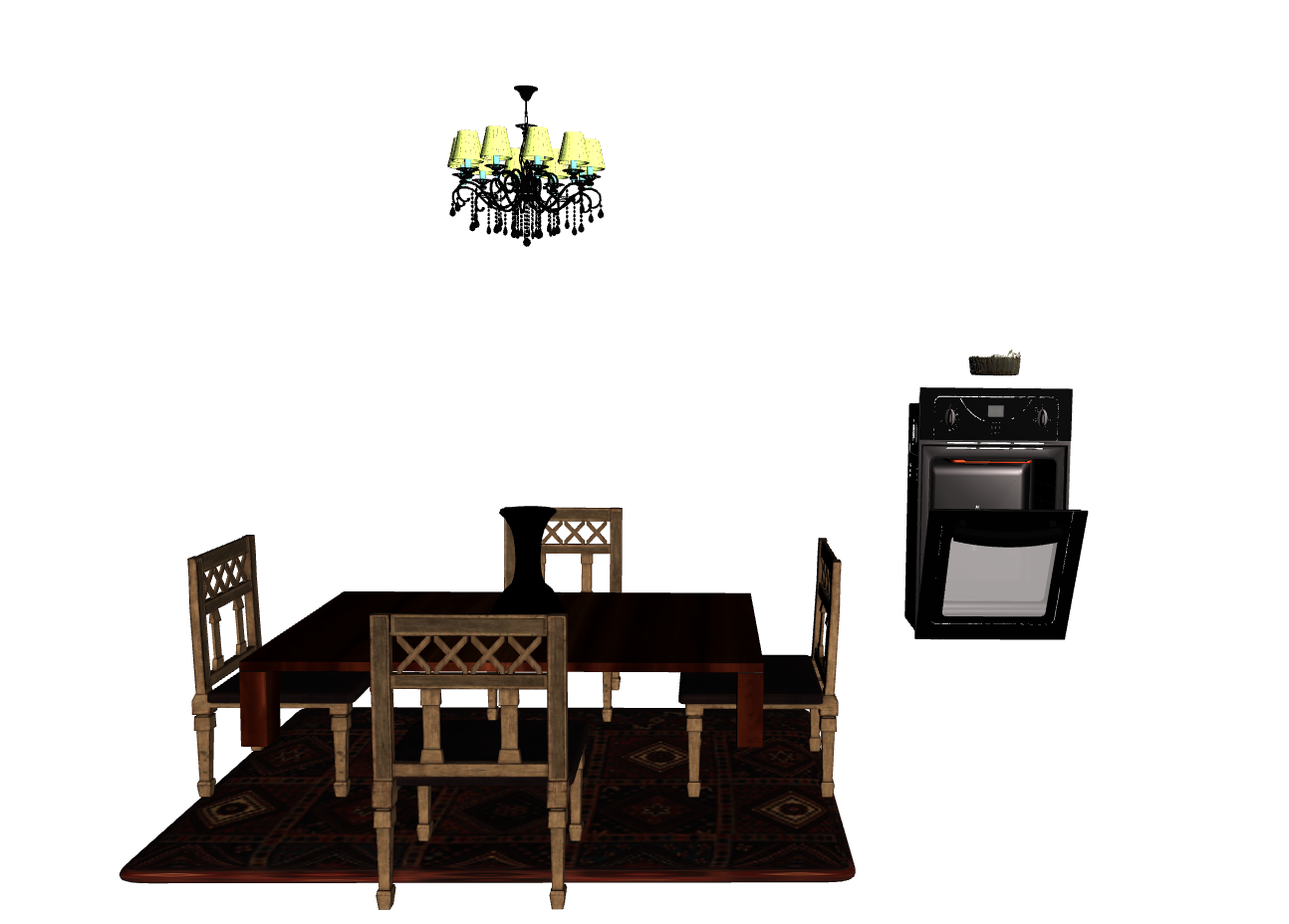}
  \end{minipage}
  
  \caption{Scene generated by the prompt \textit{"A warm and rustic kitchen with a chandelier hanging above a polished wooden dining table at the center. The table, resting on a cozy rug, is neatly surrounded by four chairs and features a simple vase as its centerpiece. In the back corner is an open oven with a freshly baked apple pie on top"}.}
  \label{fig:trimesh-render}
  \vspace{-6mm}
\end{figure}

\subsubsection{Layout Improvement}
\label{sec:layout-improvement}
Once we have the position, size, and orientation for each object we can create a simplified rendering of bounding boxes for each asset through Matplotlib \cite{Hunter:2007}. Following \cite{Yang2023SetofMarkPU} we create a rendering that adds a set of visual marks (an overlayed number) over semantically meaningful regions, which in our case are the distinct bounding boxes. This form of Set-of-Mark visual pixel prompting has been effective in grounding multimodal models and improving their effectiveness on a variety of tasks. We present this graphical 2D rendering as well as a textual state of the current scene along with the original description of the scene  as inputs back into the model to improve the layout. This form of intermediate computation mirrors that of \cite{Nye2021ShowYW} which introduces a "scratchpad" 
to explicitly write out computation steps that is particularly helpful for multi-step reasoning tasks. We continue this feedback loop until the system concludes that no changes need to be made or for a maximum number of iterations (which we set to 3 in our experiments). 

Once the positions for all the assets are set, the rotations of the assets have to be decided. Given the fact that there is no golden standard for default rotations of arbitrary 3D meshes available in most datasets, this is a very challenging problem. 
To achieve this, we first render each asset individually using Trimesh \cite{trimesh}, and then prompt a VLM to determine the current facing direction of the asset. Based on this, we determine how to rotate the asset to ensure proper alignment with its intended orientation. For simplicity, we only consider 90 degree rotations along the vertical axis (y in Unity). In our metaprompt we describe common sense forward facing directions of some everyday objects, and then prompt separately for the proposed rotation for each object in the list of assets we are adding to the scene. For details, see the Supplementary Material. 
An example of this approach is demonstrated in Figure \ref{fig:trimesh-render}, where the orientation of a chair is automatically detected, and multiple copies are placed around a table, all facing inward. Finally, the proposed updated layout, consisting of new positions and rotations of all the assets gets sent to Unity from the Flask server. At this point we can make the bounding wireframe boxes change to green and then disappear to signify that the automatically proposed placements are finished.
\begin{figure}[ht]
    \centering
    \includegraphics[width=1\linewidth]{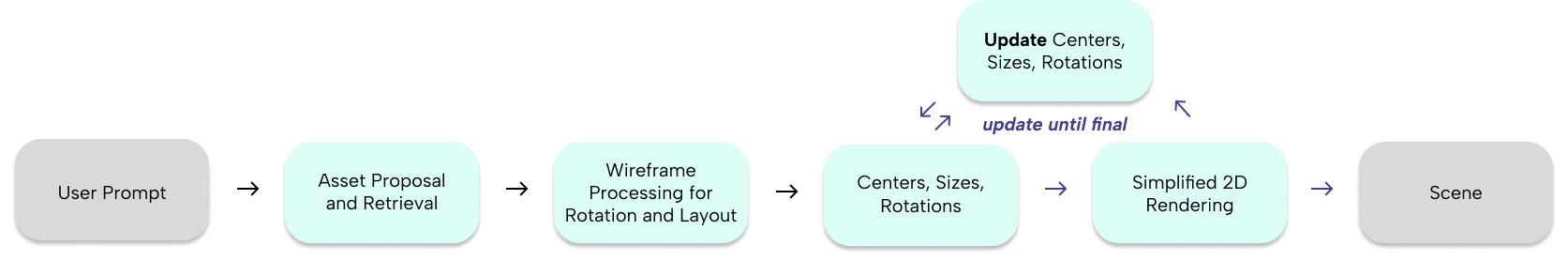}
    \caption{Overall scene layout pipeline.}
    \label{fig:scene-diagram}
\end{figure}

\subsection{Tools and Interaction}
\label{sec:tools}

\subsubsection{VR-Centered Tools and Interaction}
\begin{figure}[ht]
  \centering
  \begin{subfigure}[b]{.48\linewidth}
    \centering
    \includegraphics[width=\linewidth]{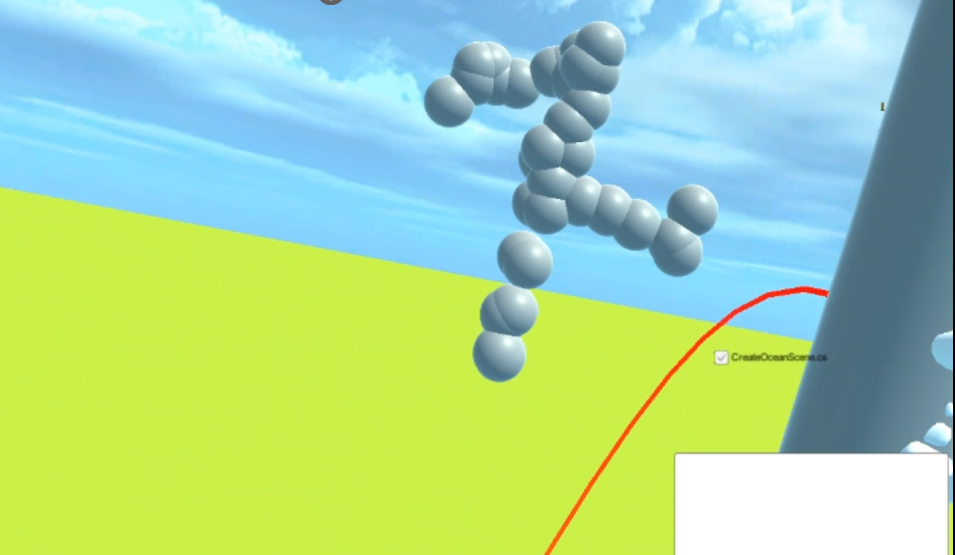}
    \caption{Tool generated for the prompt \textit{"create a wand tool that spawns spheres with a trigger press"}.}
  \end{subfigure}
  \hfill
  \begin{subfigure}[b]{.48\linewidth}
    \centering
    \includegraphics[width=\linewidth]{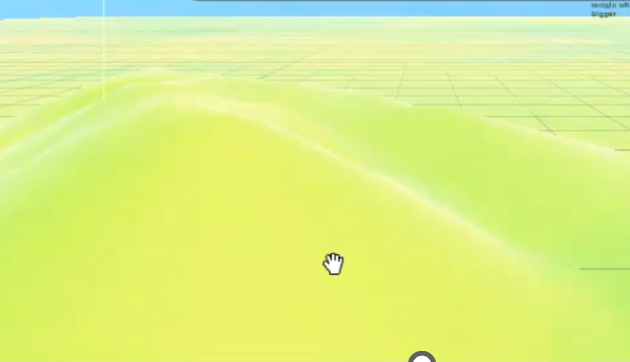}
    \caption{Aftermath of a terrain after using a tool created by the prompt \textit{"create a tool that adds hills with the trigger"}.}
  \end{subfigure}
 \centering
  \begin{subfigure}[b]{.48\linewidth}
    \centering
    \includegraphics[width=\linewidth]{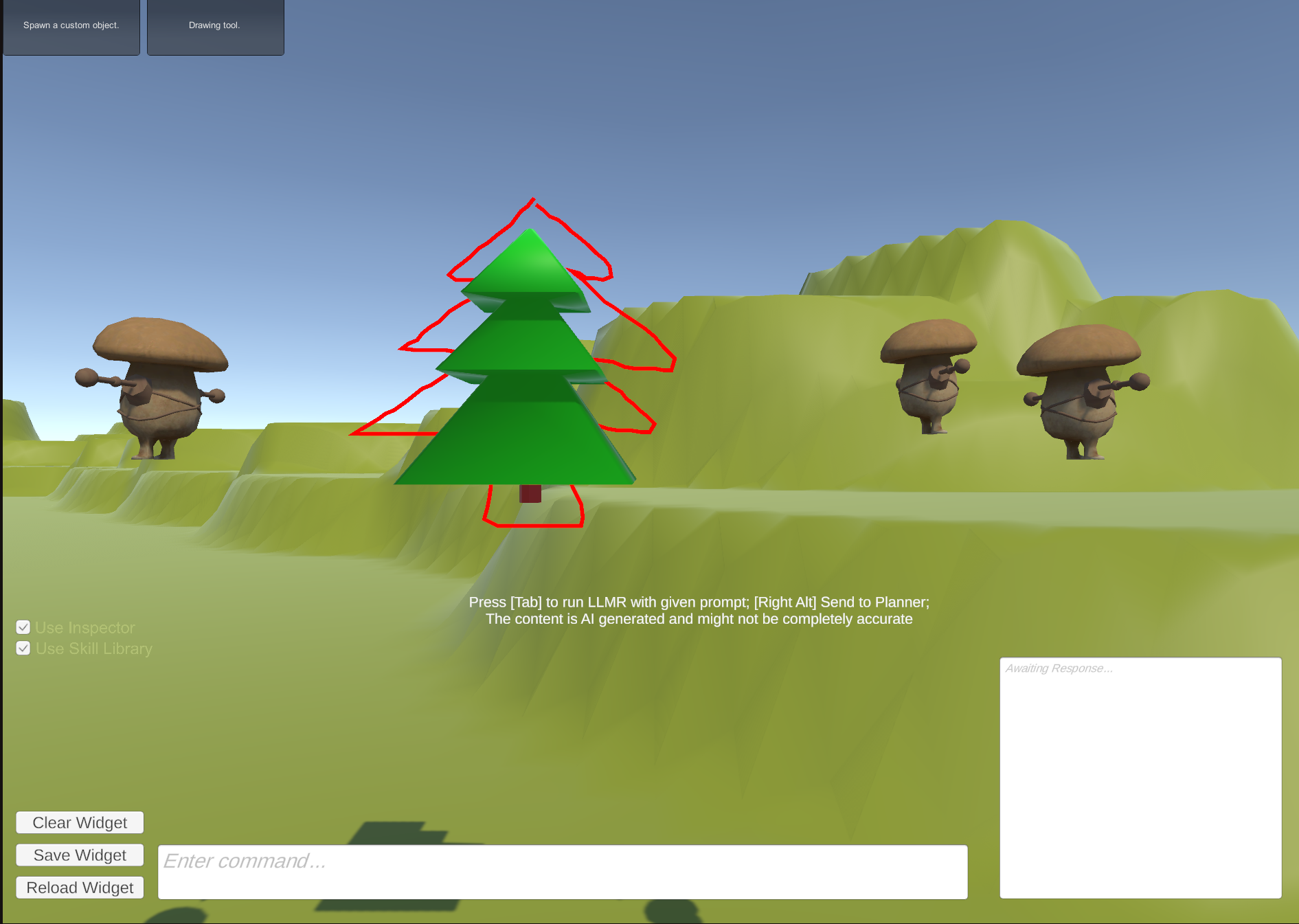}
    \caption{Going from a pine tree sketch generated by the drawing tool to a 3D object in a multi-user setting.}
  \end{subfigure}
  \hfill
  \begin{subfigure}[b]{.48\linewidth}
    \centering
    \includegraphics[width=\linewidth]{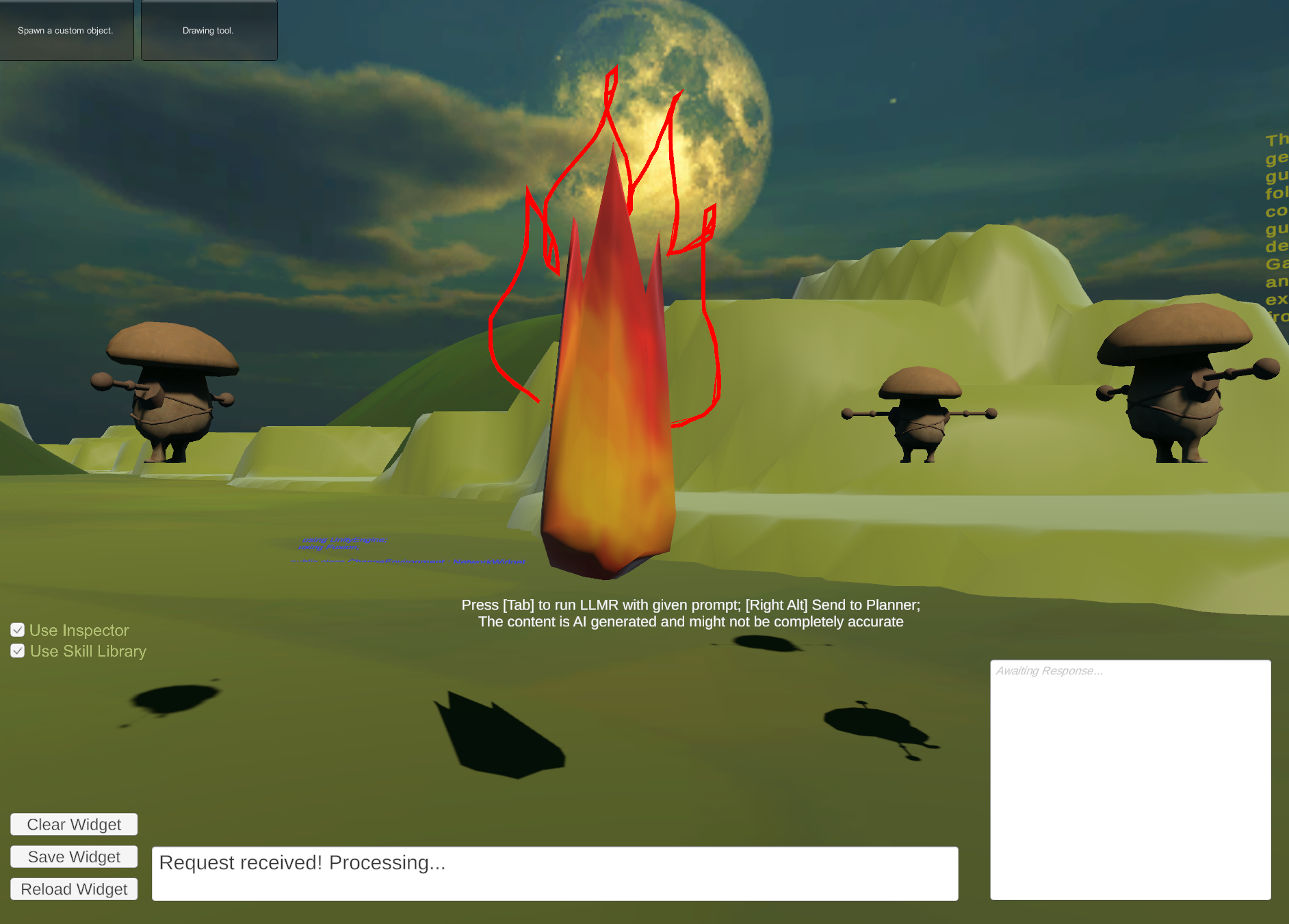}
    \caption{Going from a fire sketch generated by the drawing tool to a 3D object in a multi-user setting.}
  \end{subfigure}
  
  \caption{Examples of tools and tool output enabled by our system.}
  \label{fig:gen-tool-examples}
  \vspace{-6mm}
\end{figure}

To provide users with additional agency over how they can interact with and modify scenes, we integrated VR-centered tools and interaction into our system. We use Unity's XR Core Utilities \cite{XRCoreUtilities} and XR Interaction Toolkit \cite{XRInteractionToolkit} packages to add support for VR, including head and hand tracking. Additionally, we incorporate logic into our metaprompts to describe how to generate code using the XR Interaction Toolkit, see details in the Supplementary Material. This allows users to prompt tools into existence without overtly technical language or specific library references evident in previous work \cite{de2024llmr}. For example, a user could prompt \textit{"create a wand tool that spawns spheres with a trigger press"} or \textit{"create a tool that adds hills with the trigger"} and interact with the tool generated in their VR environment (See Figure ~\ref{fig:gen-tool-examples}). In addition to enabling tool creation, our system includes a number of default tools accessible through a graphical user interface that allow users to modify their scenes in VR more easily. These include VR tools for moving and rescaling objects, swapping a loaded asset with a different object, and drawing in 3D. 

By exposing our Coder to logic for generating code with the XR Interaction Toolkit, users can also prompt VR centered interaction into existence. For instance, a user can prompt a table and mug into a scene, realize that the mug cannot be picked up by the controller, and then prompt to \textit{"make the mug grabbable"}. Subsequently, the user is able to pick up the mug with their controller. 

Through this combination of default tools and user generated tools, users have an additional layer of control in VR over modifying the content within their scenes. In doing so, our system can account not only for how scenes are generated, but also the embodied experience of the users immersed within them.

\subsubsection{Sketch-To-Object Prompting and Multi-Modal Tools}

To allow users additional flexibility in multi-modal input, we expose our Coder to few-shot sketching tool examples, which can be prompted by users after. Rosenberg and colleagues \cite{perlin2024draw} discuss creative affordances and opportunities in sketching and speaking tools in generating immersive stories in non-3D environments. In \textit{Social Conjurer}, however, we aim to prompt for these multi-modal interfaces and augment them with multi-user networking capabilities in immersive VR and 3D, making the runtime environments better equipped to serve various use-case scenarios.

\begin{figure}
    \centering
    \includegraphics[width=1\linewidth]{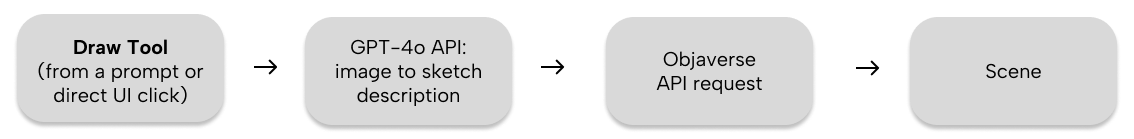}
    \caption{Draw Tool System Overview in Social Conjurer.}
    \label{fig:draw-tool}
\end{figure}

At a high-level, we aim to go from a user prompting a tool that they then use to sketch with in the 3D environment. The system then takes a screenshot of the sketch, parses it with a VLM, and returns a 3D asset (See Figure \ref{fig:draw-tool}). This involves parsing the image to a text tag, searching for that tag in the Objaverse \cite{objaverse} system, and returning a 3D asset back into the Unity scene. In multi-user environments, the synchronization of the tool, sketch, and returned object undergo additional steps that are explained in detail below.

\subsection{Multi-User Worldbuilding}
\label{sec:networking}

The purpose of \textit{Social Conjurer} is to allow for multi-user, spontaneous world creation in 3D environments, which can later be accessed through a VR headset. State-of-the-art networking libraries \cite{unity-multiplayer-networking, ai-networking, photon-engine} defined for the Unity game engine \cite{unity} offer limited support for real-time creation and synchronization of 3D assets, e.g. objects, scenes, and scripted behaviors. Networked assets instantiated through these libraries must be defined before running any given Unity project, and spawning custom assets is only implemented in primitive state through the Fusion 2 library \cite{photon-fusion}.

\begin{figure}
    \centering
    \pdfstringdef{\alttext}{This diagram depicts the architecture of a networking module for a collaborative virtual reality system. It involves a 'User Prompt' initiating a process where the 'Decider' interacts with a 'Networking Module' that includes Photon Fusion 2 and a Flask Server. The Networking Module then interacts with the 'Scene'. Four key processes are illustrated: (1) Runtime Prefab Registration, (2) Dynamic Mesh Update, (3) Assigning Scripted Behaviors, and (4) Client Synchronization. These processes enable dynamic updates to the scene and synchronization across clients in real-time.}
    \includegraphics[width=1\linewidth]{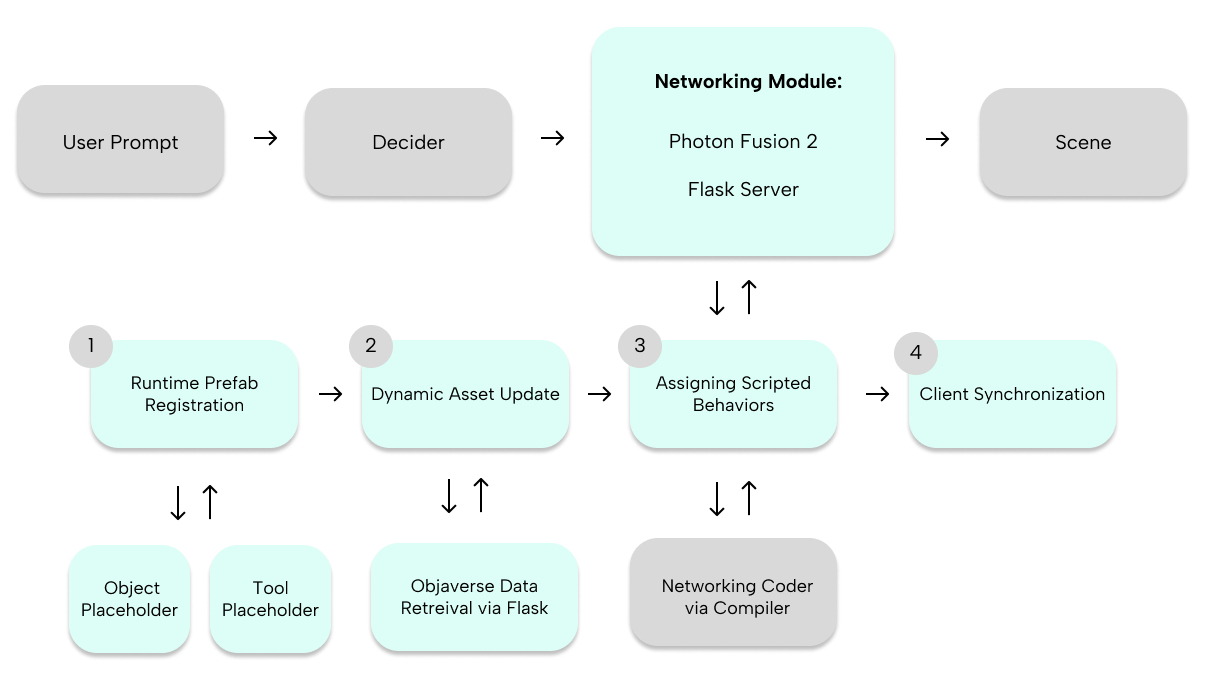}
    \caption{Networking Module for the \textit{Social Conjurer} System.}
    \label{fig:networking-module}
\end{figure}

In other words, to support \textit{Social Conjurer's} main capability of creating assets at runtime, we must modify existing networking tools and libraries to allow multiple users to generate environments together. In doing so, we augment the current system with the \textit{Networking} module, which allows the \textit{Conjurer}'s Coder and Compiler modules to spawn scripted objects with default network behaviors. Inside the Networking system, we use Photon Fusion 2 to support basic networking capabilities (creating empty shared environments and simple network communication in Unity), as well as custom Python Flask server logic to support us in transferring complex meshes and network behaviors between multiple clients. Refer to Figure \ref{fig:networking-module} for the high-level architecture overview.

\subsubsection{Networking Module}

The Networking module includes updates to the Builder (which we renamed into Coder for our system, as per discussion above), Inspector, Compiler, and Widget components from the open-sourced LLMR system \cite{de2024llmr} that we base parts of \textit{Social Conjurer} on. In particular, the Coder and Inspector modules need metaprompts that include networked properties when writing new scripts, utilizing library calls to Photon Fusion 2 functions. See the Supplementary Material for the detailed metaprompts. The strategy we take in our approach is to ensure that any script created by the Coder references Photon Fusion 2 assemblies, which if compiled successfully, will get attached to the Compiler game object in the scene. In our system, the Compiler object is promoted to a registered networked asset, so that it can serve as a proxy to any networked Coder scripts attaching behaviors to other networked objects in the scene (since all of the generated scripts are attached to the Compiler as a proxy object). 

\subsubsection{Photon Fusion 2}

Calls to the Photon Fusion 2 library happen every time the Networking Coder requests a new object, or attaches a behavior to an existing networked object. Similar to instantiating primitive objects at runtime, we override the logic to create a prefab placeholder every time an object gets instantiated with a custom prefab flag, and continue to modify that object at runtime by making calls to the Flask server that stores the object's mesh, animations and location data from Objaverse \cite{objaverse}.

A separate prefab flag is used for prompting virtual tools. When a tool is initialized, e.g. a drawing tool, an object placeholder gets created first. Then, a separate script is written by the Networking Coder to attach a tool behavior to the networked tool object. This script waits until the tool prefab gets created and registers in the network, and only afterwards does it get to assign any behaviors (e.g. drawing functionality) to the networked object. Any updates to the tool behaviors are completed with Remote Procedure Calls (RPCs) that are synchronized over the network and are communicated between clients.

\subsubsection{Python Flask Server}

Building on top of the Flask server that supports Layout Arrangement and Improvement modules (\ref{sec:initial-layout} and \ref{sec:layout-improvement}), we implement custom methods on the server to store and return the requested object's information, including the asset name, a downloadable URL of the asset, and any location information if the asset gets any spatial properties. Calls to the Flask server are completed inside of the custom Photon Fusion 2 module, right when the networked object placeholder gets initialized and updated in runtime.

\section{Evaluation}
To evaluate our system, we conducted an in-person user study with 12 participants (6 pairs). Addressing our second and third research questions, our objectives of this study were to (1) assess the usability and user experience of our system, (2) understand how social contexts shape Human-AI co-creation of virtual worlds, and (3) use our system as a design probe for gathering perspectives on social applications of prompt-based 3D co-creation. Our institution's research and compliance committee approved all of our procedures, outlined in this section. We supplement this study with an evaluation of our system's spatial reasoning. 

\subsection{Participants}
Twelve participants were recruited from a company in the United States to take part in our study. Participants (6 females, 6 males) were recruited through group chats and word of mouth. Five participants were between the ages of 18-24, and seven participants were between the ages of 25-34. Most participants were familiar with GenAI. Based on reporting, 8\% were very unfamiliar with GenAI, 33\% were slightly familiar, 33\% were familiar, and 25\% were very familiar. Familiarity with augmented, mixed, or virtual reality was more varied, as 8\% were very unfamiliar, 17\% were unfamiliar, 8\% were slightly unfamiliar, 8\% were neutral, 42\% were slightly familiar, and 17\% were familiar. Our study was conducted in pairs, of which 58\% of participants reported having interacted with their partner prior to the study. Participants filled out their consent form within a week prior to their study session. P1 and P2 engaged in a pilot version of the study. After running the pilot, we decided to incorporate a more structured tutorial of our system into our main study procedure. 

\subsection{Procedure}

\begin{figure}
    \centering
    \includegraphics[width=1\linewidth]{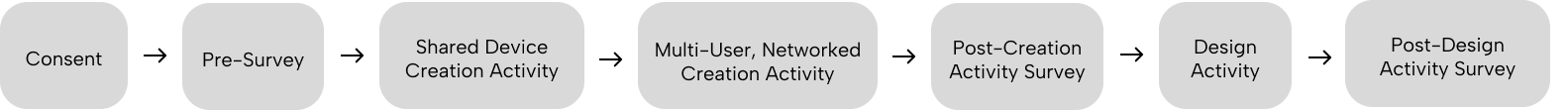}
    \caption{An overview of the study procedures.}
    \label{fig:user-study}
\end{figure}

Our in-person study consisted of a survey, scene creation activities where participants interacted with our system, and a design task. Please refer to Figure \ref{fig:user-study} for the overview of the procedure. 

Participants first completed a pre-survey to provide background information. Next, they were introduced to the \textit{Social Conjurer} system, receiving a brief overview of the interface and instructions on navigating and modifying their scene. Afterward, participants engaged in two collaborative, open-ended scene creation tasks, each using a different version of the system (described below) and collaboration style. While both versions supported prompt-based scene creation, each was designed to test distinct features, introduced separately to manage participants’ cognitive load.

\subsubsection{Shared Device Creation Activity}
The first task involved building a scene on a shared device with one laptop and one Meta Quest 3 VR headset. Participants were instructed to create an outdoor virtual scene together using the prompt-based interface. Although the prompting textbox was only accessible on a desktop, participants were provided with the option of viewing, navigating, and modifying their scenes in VR through controller-based interaction. Participants wore a hands-only avatar that was responsive to the VR controller input and hand movements. Embodied interaction, in how VR allowed participants to engage more directly with the virtual world, and spatial reasoning, in using our layout arrangement and improvement modules, were core focus areas in this system version. 

\subsubsection{Multi-User, Networked Creation Activity}
In this second task, participants were instructed to \textit{"collaborate on creating a 3D scene with your partner"} in a networked virtual environment, using separate devices (two laptops and one Meta Quest 3 VR headset). Each participant interacted with the shared scene through their individually assigned device. Participants were instructed to generate a scene together using language-based prompts and sketches. Due to the nature of the networked environment, changes made by one participant were viewable on the other's device and vice versa. Hence, the core focus areas for the second version of the system involved networking virtual environments and the additional use of sketches as a prompt input. During this task, participants viewed default, humanoid mushroom avatars from a third-person perspective, controllable with keyboard input. However, they were provided with the option to view their scene in VR using the provided headset. Each creation task lasted 10 to 15 minutes, until participants were satisfied with the scene they created, time ran out, or system errors prevented further creation of scenes. Following these creation tasks, participants complete a post-creation activity survey about their experience.

\subsubsection{Design Activity}
Next, participants engaged in an individual design task where they were instructed to draw a concept sketch representing their vision of social applications for a future version of this system. They were prompted to \textit{"consider how multiple people might interact with the system together, and what makes these interactions meaningful or fun."} Participants were provided with paper and drawing utensils and had around seven minutes to sketch their idea(s). Afterward, participants shared their ideas out loud and were asked about their experience across the two collaboration activities. Lastly, they completed one more question in their survey, which revolved around brainstorming their vision's positive and negative implications.

\subsection{Data Collection and Analysis}
To analyze the qualitative data, we used data triangulation \cite{fielding2012triangulation} to identify emergent themes relevant to our research questions. This process involved a collaborative and iterative coding process. Two authors employed an inductive approach to uncover an initial sets of themes in each data type. These categories were then discussed, triangulated, and refined between the authors to pinpoint the topics most relevant to our research questions. Following the guidelines for qualitative HCI Practice by McDonald and colleagues \cite{mcdonald2019reliability}, we identified recurring topics pertinent to our study and linked them to form broader themes. According to these guidelines, the themes we developed did not have to \textit{"align with the most prevalent set of codes but instead those that are salient to the research question or inquiry." }\cite{mcdonald2019reliability}. We use descriptive statistics to analyze quantitative data. The types of data collected as a part of this analysis were as follows:

\subsubsection{Pre- and Post- Activity Survey}
All participants completed survey items (See Supplementary Materials) both before and after the study tasks. The survey gathered data on demographic information, participants' prior experience with GenAI and immersive technologies, and whether they had previously interacted with their study partner. Additionally, it included several questions designed to assess user experience with the system and participants' perspectives on its potential. To evaluate the networked version of the system, we included a system usability measure consisting of 10 items on a 7-point Likert scale (1 = Strongly disagree, 7 = Strongly agree). These items were both adapted from previous studies \cite{de2024llmr} and newly created to address our specific research questions. In a short-answer section of the survey, participants were asked: \textit{"In what scenarios would you find a tool like the one used in this session valuable?"}, \textit{"Are there any situations where you would prefer not to use this tool? If so, why?"}, and \textit{"What did you like and dislike about the tool you used today?"} Following the design activity, participants responded to an additional short-answer question: \textit{"Brainstorm positive and negative implications of your vision. Describe them here."}.

\subsubsection{Audio and Screen Recordings}
Each participant consented to audio and screen recordings of their study session. Audio recordings captured participants comments during their tasks and discussion surrounding their design artifact. These recordings were transcribed using Microsoft Teams transcription software. Screen recordings captured interactions with the system during both scene creation tasks. 

\subsubsection{Design Artifacts} We collected the concept sketches made by participants during the design activity.

\subsection{Evaluation of Spatial Reasoning}

To assess the spatial reasoning capabilities of our system, we developed a specialized benchmark focused on positional reasoning from textual inputs. This aligns with existing datasets which require inferring spatial relationships between entities from text \cite{Mirzaee2021SPARTQAAT, Shi2022StepGameAN, Weston2015TowardsAQ}. Unlike \cite{Mirzaee2021SPARTQAAT}, which assess tasks like finding blocks, selecting objects, or answering yes/no questions, our benchmark exclusively targets the precise understanding of spatial positions. \cite{Shi2022StepGameAN, Weston2015TowardsAQ} involve presenting longer form narratives with distractors to infer spatial relationships between objects, whereas we focus on the system's ability to propose accurate and logical candidate locations within a 3D space.

We do this by prompting GPT-4 to generate diverse scenes with explicit spatial relationships between objects. These relationships were constrained to one of six directions: left, right, behind, in front, below, and above, ensuring a straightforward parsing process. For each scene, we parsed the descriptions and generated ground-truth mappings for object relationships. To minimize parsing errors, we limited extractions strictly to relationships explicitly defined in the text, avoiding any inferred connections. For example, if object A is above object B and object C is above A, we extract only "object C above object A" without assuming "object C above object B." This approach ensures the process remains fully automated, scalable, and suitable for initial evaluation. Future iterations will extend and validate the dataset to include more complex reasoning cases. In total, we generated 75 scenes with 840 relationships. See Supplementary Material for additional information on scene prompts and spatial relations.

\section{Findings}

\begin{figure}[ht]
  \centering
  \begin{subfigure}[b]{.48\linewidth}
    \centering
    \includegraphics[width=\linewidth]{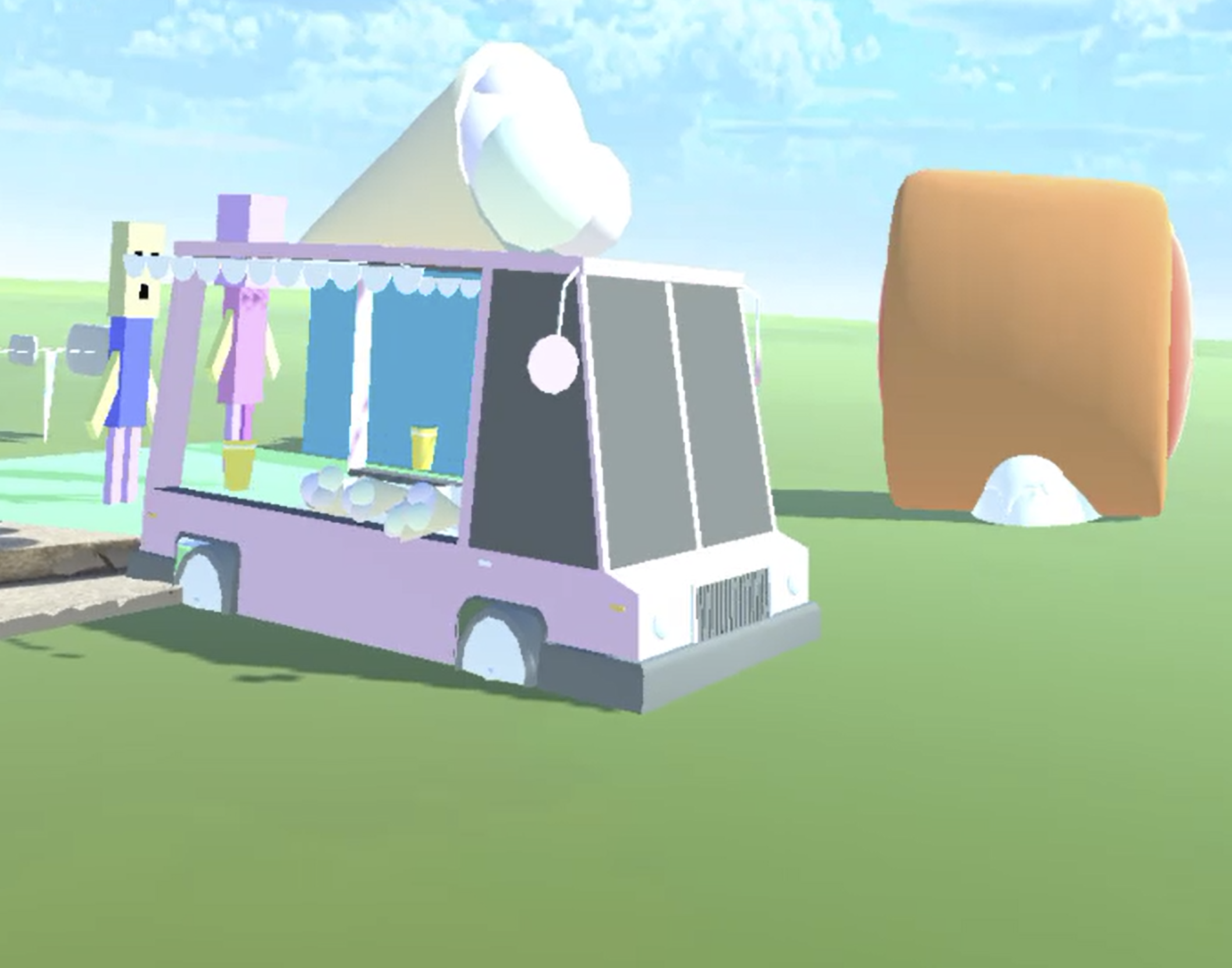}
    \caption{\textit{"Ice cream truck with a hot dog and parents in the background"} scene created by participants.}
  \end{subfigure}
  \hfill
  \begin{subfigure}[b]{.48\linewidth}
    \centering
    \includegraphics[width=\linewidth]{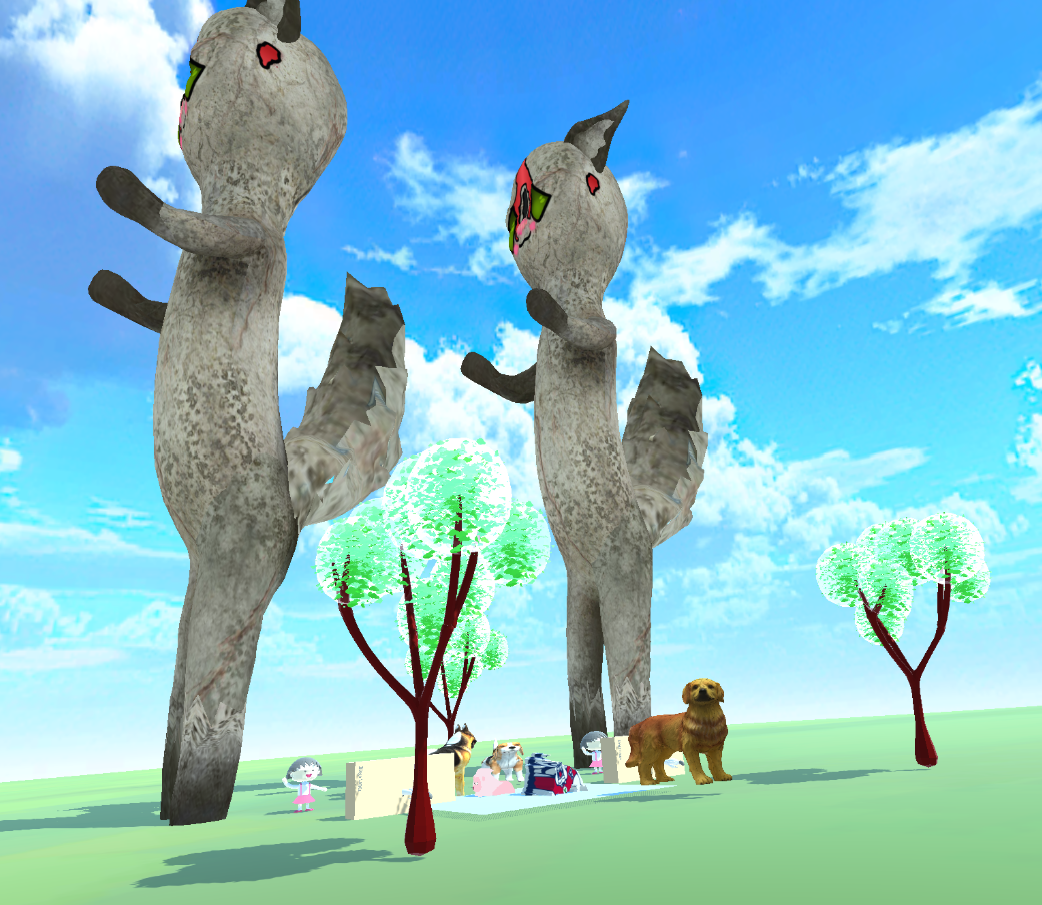}
    \caption{\textit{"Cute dog park with evil cats"} environment generated by participants.}
  \end{subfigure}
 \centering
  \begin{subfigure}[b]{.48\linewidth}
    \centering
    \includegraphics[width=\linewidth]{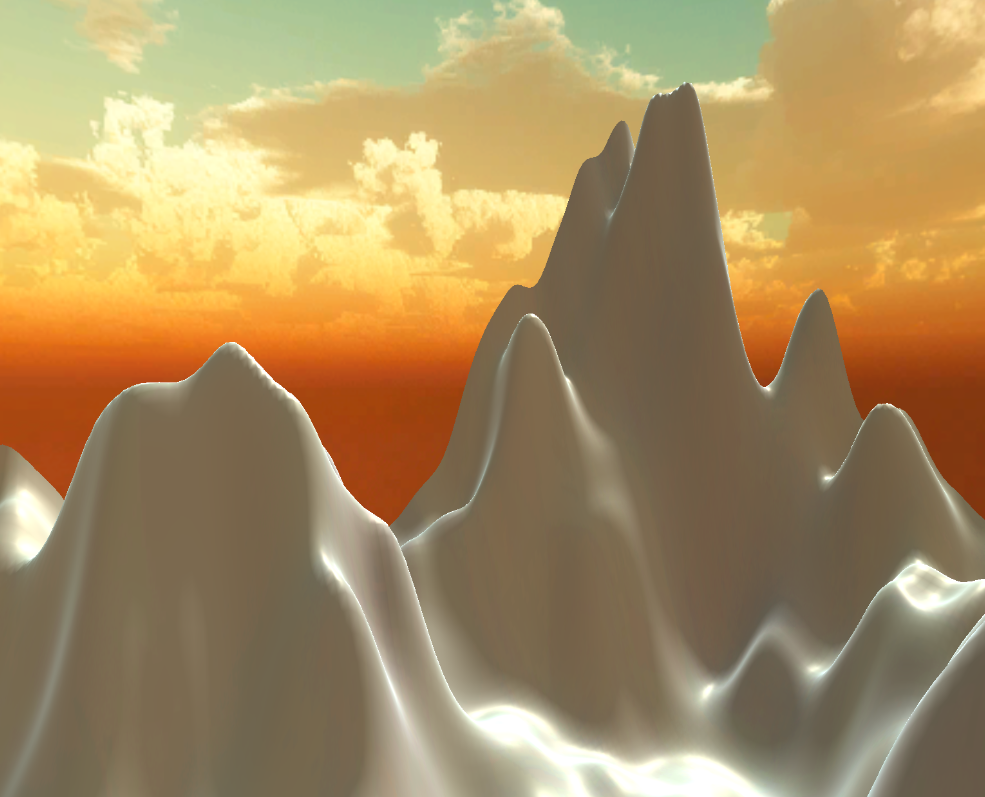}
    \caption{\textit{"Snowy Hike"} environment prompted by participants.}
  \end{subfigure}
  \hfill
  \begin{subfigure}[b]{.48\linewidth}
    \centering
    \includegraphics[width=\linewidth]{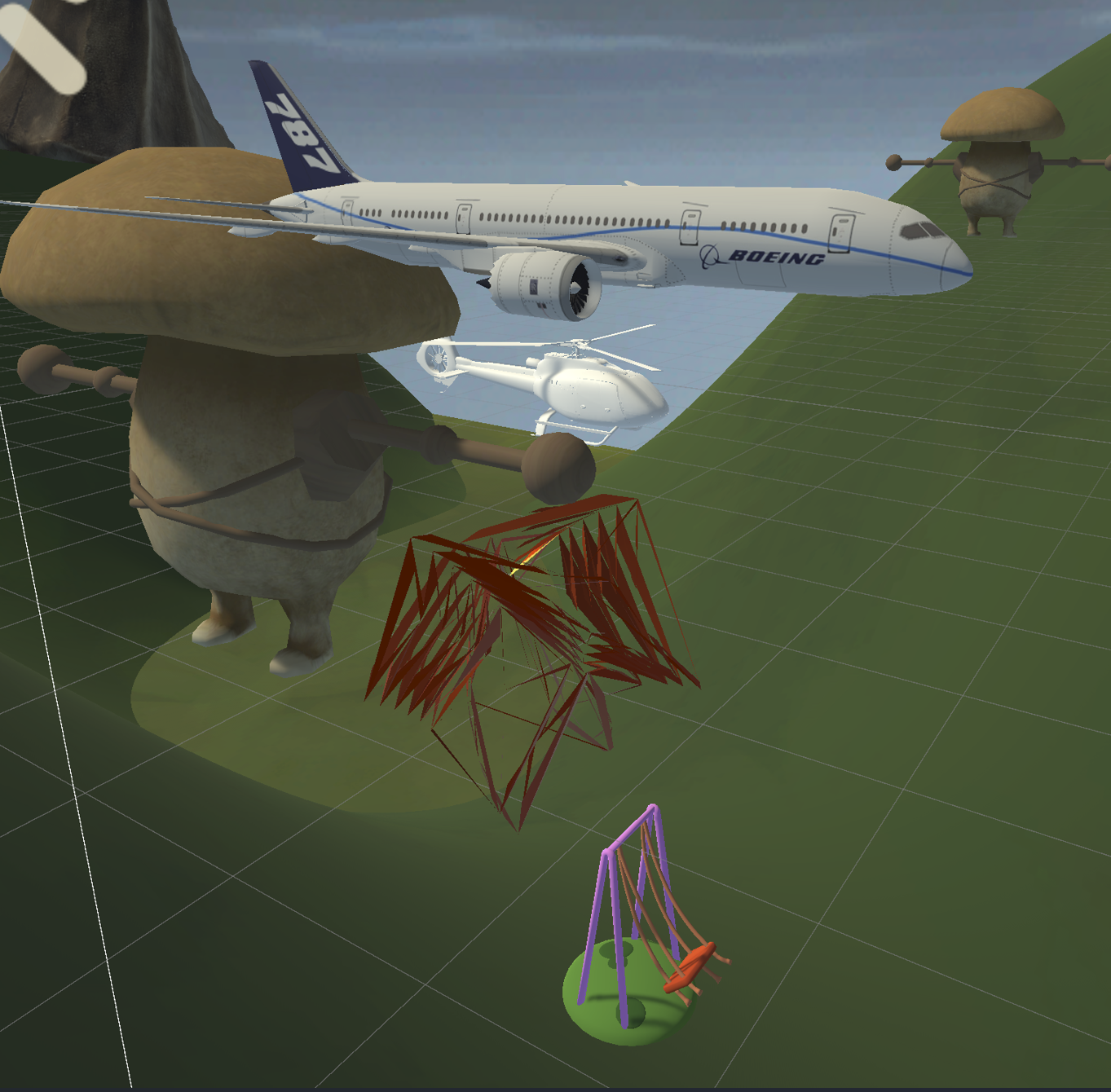}
    \caption{Multi-user environment after prompting objects from text and sketches.}
  \end{subfigure}
    \caption{Examples of scenes generated by participants in our user study.}
  \label{fig:scenes-examples}
  \vspace{-6mm}
\end{figure} 

\subsection{Qualitative User Study Findings}
Based on participants' open-ended responses to survey questions, verbal comments, prompts, and design artifacts, we identified key themes relevant to assessing how social contexts shape spontaneous world creation (RQ2) and the challenges and opportunities associated with prompt-based 3D world creation (RQ3). In this section, we outline these themes alongside areas of improvement for our system. Refer to Figure \ref{fig:scenes-examples} for some of the scene examples generated by the users.

\subsubsection{Controlled Randomness --  Discrepancies in User Intent and AI Interpretation}
There were multiple instances where participants noted a mismatch between their intentions for a scene and our system's interpretation of the scene based on their provided input. This was called out by P6 who commented in their survey, \textit{"The [tool] definitely needs improvements on understanding prompts..."} One of the noteworthy discrepancies was contextual misunderstanding, where the system would generate an object that was semantically appropriate but contextually inaccurate. For instance, when P4 prompted for a helmet to go with the bicycle already present in the scene, our system loaded a football helmet, causing him to remark \textit{"Well, it was meant to be a cycling helmet."} This mismatch was evident in aesthetic differences as well. In particular, our system had challenges fully capturing the artistic intent of users. As P5 and P6 prompted objects into their park scene on a shared device, they expressed discontentment with an asset that was a volumetric capture of real children amidst the more stylized, cartoonish assets in their scene. P5 expressed \textit{"What is going on with the children... They're kind of creepy."} leading to a subsequent prompt by P6 to \textit{"make the children look more normal."} P6 would also go on to prompt \textit{"Make everything a little bit prettier"} during the multi-device creation activity. P2 reflected on aesthetics generated by the system in their short response, stating \textit{"if I come to the system with, e.g. a generic "theme park" idea, it immediately becomes surreal and avant-garde, thanks to the system's quirks."} Additionally, discrepancies were especially prevalent during the sketch to 3D object task. Influenced by participants' individual drawing ability and stylistic approach, the model would sometimes misinterpret their sketches. 

The system's misinterpretation of prompts introduced an element of controlled randomness, shaping the trajectory of scene creation. There were a couple of instances where unintended generated assets inspired new prompts. For example, P3 mistakenly sent a scribble into our sketch to 3D object pipeline, which caused a bicycle to appear. Following this, P3 prompted for a helmet and a bike chain to pair with this generated asset. Similarly, P12 drew a sketch that was mistaken for an object of people sitting in a roller coaster cart. P12 stated, \textit{"Well now I should draw a roller coaster now that we have people for the roller coaster"}, and proceeded to draw a roller coaster as the next prompt. Hence, the element of chance involved in asset retrieval sometimes influenced participants' ideation of subsequent content. 

The (mis)alignment between user intention and AI output, or the \textit{"system's quirks"} (P2), was a significant topic in P2's short answer responses. While P2 \textit{"enjoyed the element of surprise in seeing how the system interpreted my prompts/sketches,"} he expressed interest in understanding the AI's decision-making process. P2 also called into question the extent to which this should be corrected, stating \textit{"How do we massage the large pre-trained multimodal models at the system's core to *intentionally* interpret the user's memories/dreams/ideas with an artistic intent? Should the tool always be half-broken and opaque, so as to remain interesting? Or should it be seamless, with the user able to perfectly have their design intent reflected by the system?"} Moreover, P2 grappled with how users might modify their vision to assimilate with AI's tendencies, expressing \textit{"One's understanding of their own world could shift to better align with AI-interpretable prompting techniques. This could color their own life with a kind of generic, middle-of-the-road, AI-slop type quality."} 

\subsubsection{Collaborative and Solo Dynamics in Prompt-Based Co-Creation} 
Patterns of both collaborative discovery and solo exploration of prompt-based creation were observed during the participants' creation tasks. In the shared device creation task (the social context of this format, where participants were seated near each other), enabled them to work together to determine what content to prompt for (e.g., \textit{"Wanna do like a pet park?"} - P10), but also how to craft a language-based prompt for a spatial environment (e.g., \textit{"Can we say add snow on the top?"} - P6). As content populated into the scene, we also observed participants work together to locate where the content appeared and debrief together on the accuracy of whether it aligned with the idea they had in mind. Although most participants did not take advantage of the ability to navigate their scenes in VR, those who did experienced a more embodied version of presence within their generated scenes that, in turn, facilitated the process of discovery. After prompting for a \textit{"hilly region with a cliff"}, P3 and P4 took turns exploring their co-created mountainous scene in VR as the other observed their point of view and guided them through their scene, fully viewable on the desktop. Hence, they experienced an embodied form of collaborative exploration to engage fully with their Human-AI generated scene. In the subsequent creation task, where devices became localized to each participant, the discovery process often became more of a solo endeavor despite the networked environment. When asked about their perspectives on the two collaboration styles, multiple participants (P3, P4, P5, P7, P8) noted that they spoke to each other less during the networked creation activity than the shared device activity. While describing why the shared device activity was more enjoyable to P7, she mentioned \textit{"I think it's because you're both looking at the same thing and you could see what was going on and like you know, we were both reacting to it and talking about it. Whereas in the second one I couldn't really see what he was creating, or at least like I wasn't sure and I was figuring it out on my own"} and later remarked, \textit{"We need some way to communicate outside of the tool"}. Her partner, P8, responded by describing how cognitive load got in the way of communication while on separate devices. This communication breakdown meant less scene planning discussion between pairs during this activity. Compared to the shared device activity, this appeared to correlate to less thematically cohesive scenes generated during the multi-device activity. However, this was not always the case, as P11 and P12 verbally decided a direction for their scene at the start of the activity. Furthermore, several participants noted the value of networked environments in remote working scenarios (P8, P9).

\subsubsection{Perspectives on Future Applications of Social Conjurer}
\begin{figure}[h]
    \centering
    \begin{subfigure}[b]{0.45\textwidth}
        \centering
        \includegraphics[width=\textwidth]{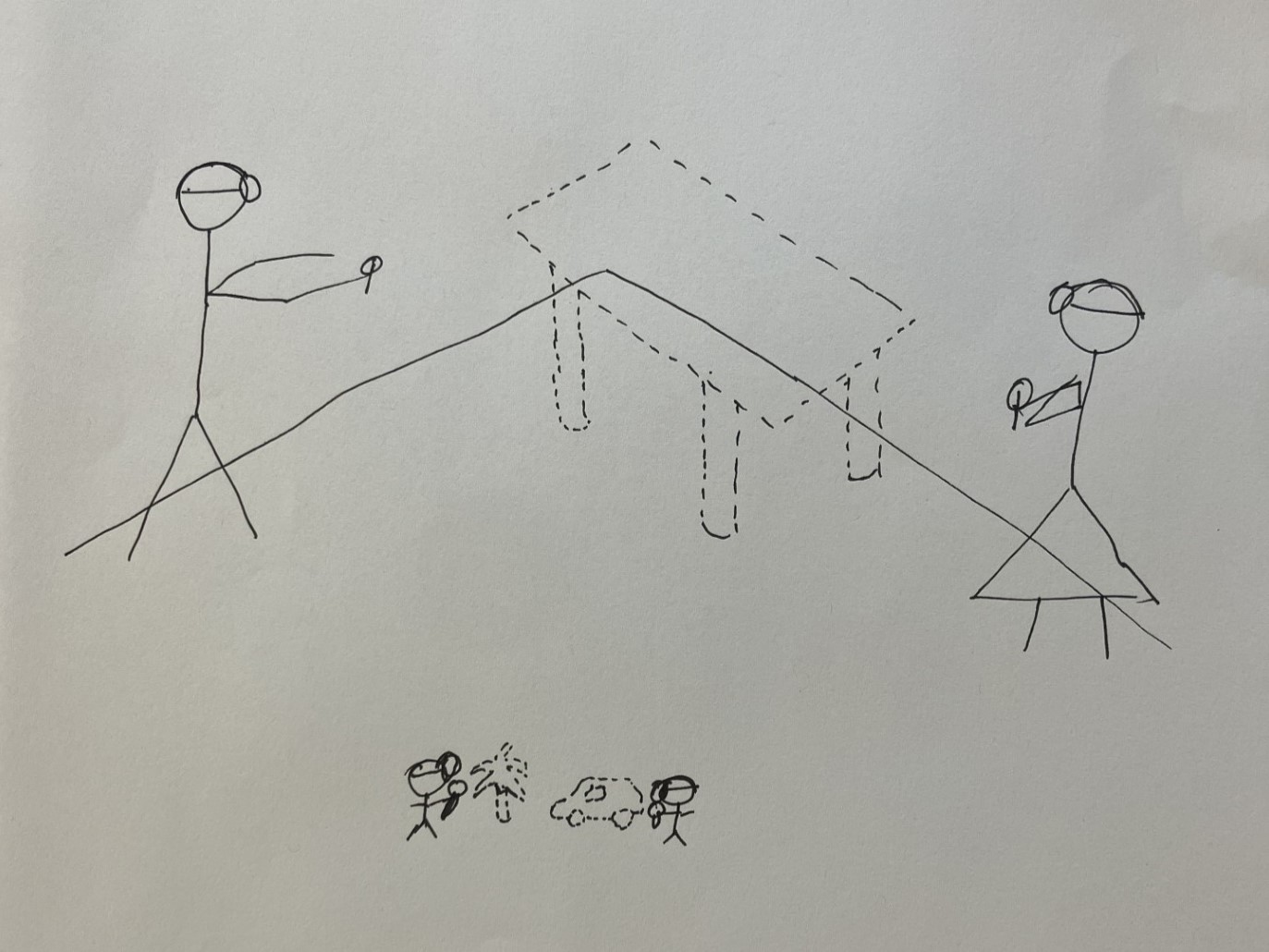}
        \caption{Concept sketch by P4}
    \end{subfigure}
    \hfill
    \begin{subfigure}[b]{0.45\textwidth}
        \centering
        \includegraphics[width=\textwidth]{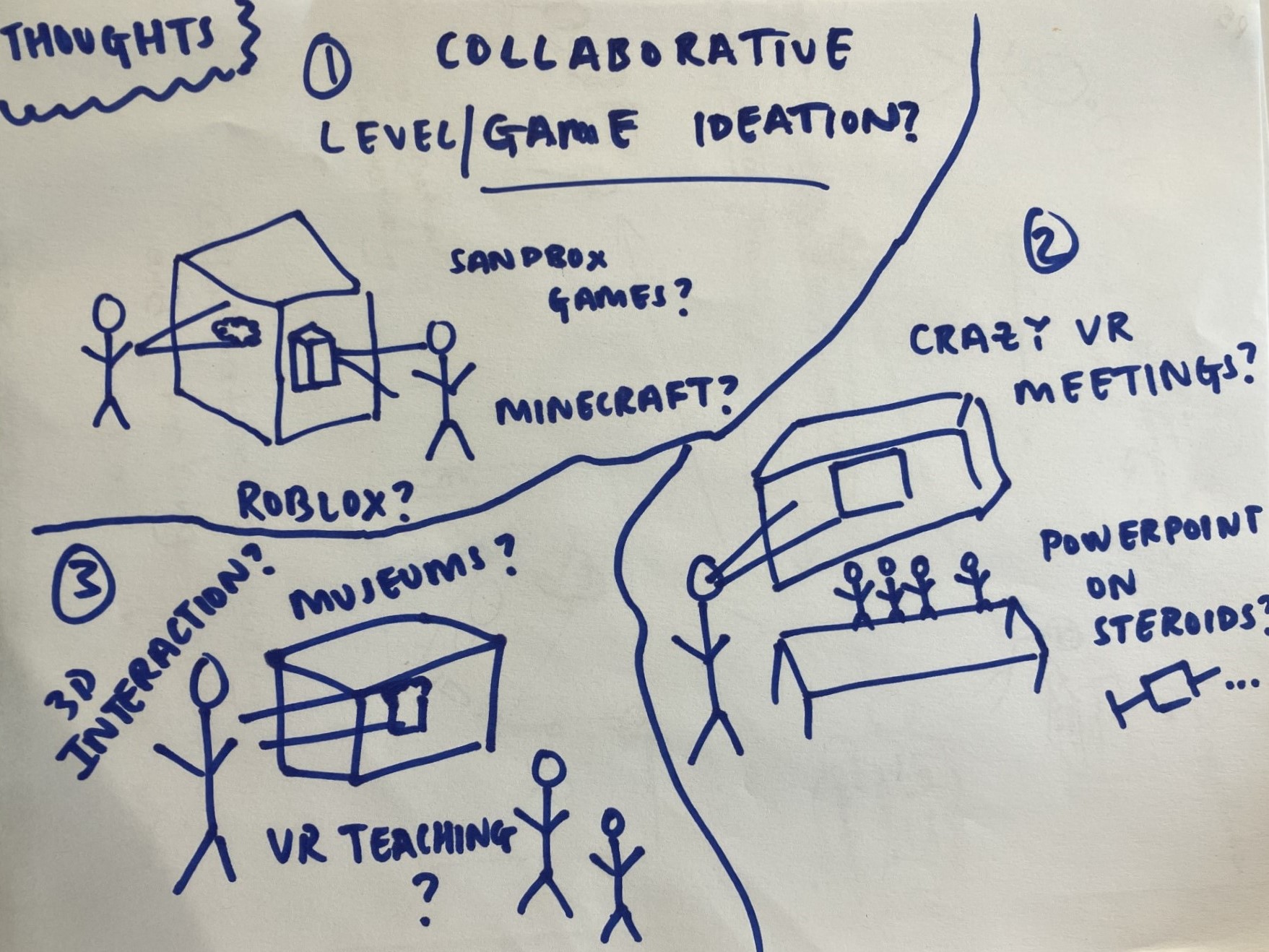}
        \caption{Concept sketch by P5}
    \end{subfigure}
    \caption{Examples of concept sketches drawn by participants during the design activity where they were instructed to sketch ideas for social applications of a future version of the \textit{Social Conjurer} system}
    \label{fig:design-examples}
\end{figure}
Based on concept sketches made by participants during the design task (See Figure ~\ref{fig:design-examples}) and open-ended responses, there were a number of reoccurring topics in regards to potential social applications for future iterations of \textit{Social Conjurer} (See Table ~\ref{tab:design-ideas-tab}). Among the most prevalent included collaborative layout design (P3, P4, P7, P11. P12) and workplace collaboration (P5, P7, P10). For example, P10 described the potential of using a system like this to create personalized virtual workspaces for remote meetings, and P5 described its potential for generating \textit{"powerpoint on steroids"} where  assets are generated as people speak during meetings. Learning was also an application of interest to participants. P5's second vision centered on the classroom, where a teacher discusses a historical  topic and \textit{"suddenly it appears"} in front of students. P8 described how spatial awareness could be leveraged to create snapshots of soccer game plays to help train athletes to improve their decision-making on the field. P3, P6, and P9 expressed the potential of prompting inclusion into existence. While referencing the popular Roblox game, \textit{Dress to Impress}, where users dress their avatars to fit certain themes, P3 noted the lack of great masculine clothing options. Consequently, he saw potential in using a system like this to fill this non-inclusive gap, allowing users to be \textit{"untethered"} from game designers' decisions. P6 noted accessibility in her idea, describing a vision where users could prompt disability-friendly architecture into existence.  
\begin{table}[]
\begin{tabular}{|p{0.15\linewidth}|p{0.8\linewidth}|}
\hline
\rowcolor[HTML]{000000} 
{\color[HTML]{FFFFFF} \textbf{Participant   ID}} & {\color[HTML]{FFFFFF} \textbf{Design Idea(s)}}                                                                          \\ \hline
1                                                & Creating a Next-Level Minecraft-like Game \\ \hline
2                                                & Embedding Social Context and   Networks into the Prompt-based Scene Generation Pipeline; Sharing Memories through Games                                                                               \\ \hline
3                                                & Collaborative Interior Design; Prompting Additional Clothing Assets into Games                                          \\ \hline
4                                                & Collaborative Interior Design; Collaboratively Generating Education Environments                                        \\ \hline
5                                                & Collaborative Game Development; Enhanced Powerpoint Presentations; Supplementing Classroom Learning Experiences         \\ \hline
6                                                & Collaboratively Generating Inclusive Virtual Worlds                                                                     \\ \hline
7                                                & Enhancing Whiteboarding of System Design / Generating Diagrams Based on Discussions; Collaborative Architecture Design  \\ \hline
8                                                & Generating Spatialized Soccer Replays for Athletic Training                                                                         \\ \hline
9                                                & Audio Annotation                                                                                                        \\ \hline
10                                               & Customizing Virtual Workspaces   for Remote Meetings                                                                    \\ \hline
11                                               & Collaborative Landscape Layout Planning; Collaborative Event Layout Planning                                            \\ \hline
12                                               & Collaborative Interior Design                                                                                           \\ \hline
\end{tabular}
\caption{Descriptions of participants' ideas during the design activity where they were instructed to sketch ideas for social applications of a future version of the \textit{Social Conjurer} system.}
\label{tab:design-ideas-tab}
\end{table}

\subsubsection{User Experience and Design Ideas}
Through the open-ended survey questions, participants highlighted several key aspects of the tool's functionality and user experience. Many users appreciated the tool's potential for rapid prototyping and idea generation, particularly in collaborative settings. For instance, P3 mentioned using the tool for \textit{"rapid prototyping of scenes for idea generation/ideation,"} while P12 found it useful for \textit{"trying to sketch something out together, or trying to quickly mock up a proof of concept view in 3D."} Comments about the system's user experience, with users finding the tool enjoyable and entertaining to use; P4 mentioned the \textit{"entertainment aspect to building an environment together in the tool,"} and P1 enjoyed the \textit{"element of surprise in seeing how the system interpreted my prompts/sketches."} Some users noted that they generally dislike the default Unity UI (P10), but \textit{"the tool made things more simple"} for them to navigate the system and create scenes \textit{"without much learning curve or effort"}. Feature ideas and design implications were frequently discussed, with users suggested improvements such as better understanding of prompts, more interaction between objects, and enhanced mechanics. P1 noted the need for improvements in prompt understanding, stating, \textit{"I would have liked to further explore the feedback aspect: can the system understand the changes that I've made to the scene, and how to further adapt to these?"} Accessibility was a concern for some users, with P3 questioning \textit{"how non-visual people would interact with this tool."} Independent of our system, the quality of 3D assets pulled from external web libraries through Objaverse \cite{objaverse} and the interface of Unity were also mentioned, with users noting the diversity of models and objects but also the lower quality of some models. P4 commented on the diversity, saying, \textit{"The diversity of models \& objects was also impressive, though usually the models were lower quality and wouldn't be ready for a production scene."} Lastly, the multi-user nature of object and scene ownership were briefly touched upon, with P9 suggesting  \textit{"I think there is some scope for improvement in the multi-agent collaboration scenario where we can get clarification into who created the scene, and who is drawing currently."}

\subsection{Quantitative User Study Findings}

\begin{figure}
    \centering
    \pdfstringdef{\alttext}{This figure shows boxplots of Likert-scale responses from 12 participants across several questions measuring user experience, satisfaction, and collaboration during the study. The questions focus on ease of collaboration, frustration levels, synchronization with partners, and overall satisfaction.}
    \includegraphics[width=1\linewidth]{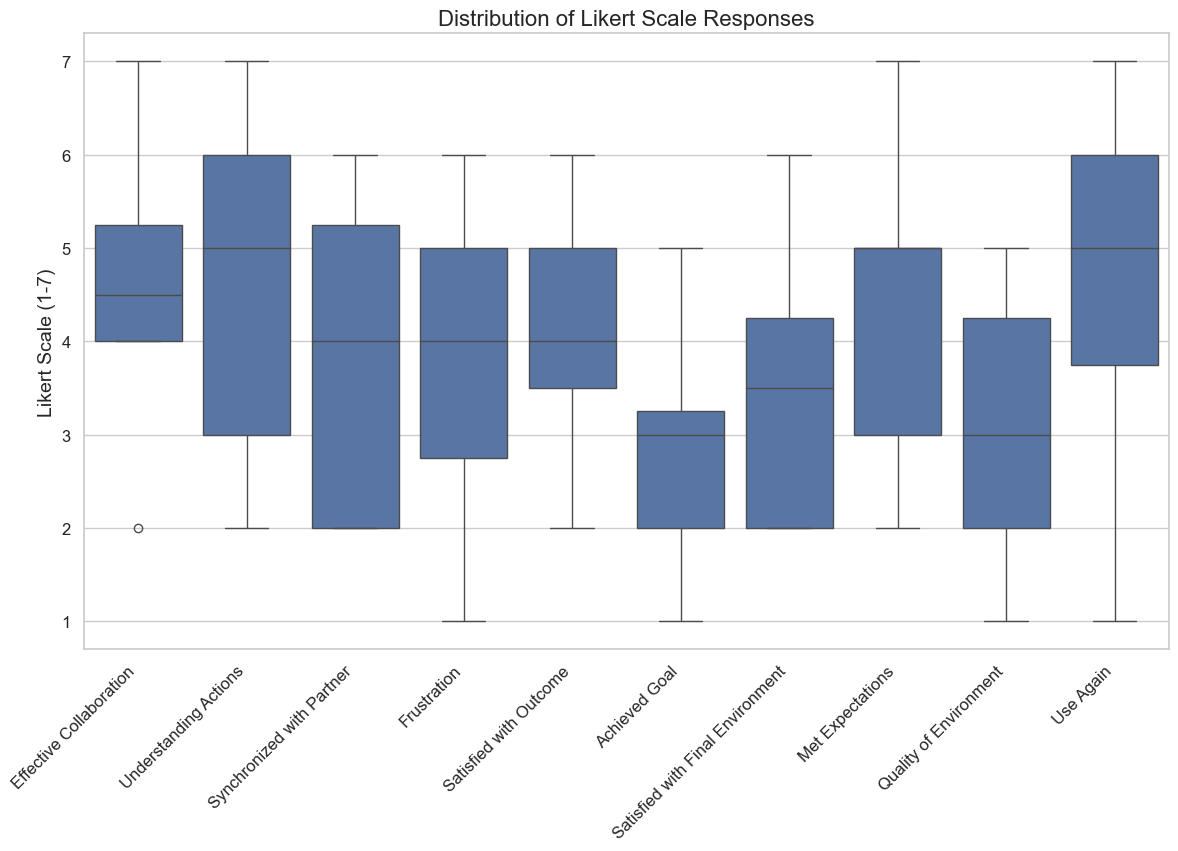}
    \caption{Distribution of 1 (Strongly Disagree) - 7 (Strongly Agree) Likert-Scale Responses for User Experience and Collaboration with the System.}
    \label{fig:likert-responses}
\end{figure}

In this section we report on quantitative findings focused on key aspects of the collaborative experience, including ease of collaboration, synchronization, user frustration, satisfaction with outcomes, and future intent to use the tool (See Figure \ref{fig:likert-responses}).

\subsubsection{Overall Collaboration Experience}

On average, participants rated the system favorably with respect to collaboration. For instance, the question \textit{"The tool facilitated effective collaboration with my partner"} had a mean score of 4.75 (\textit{SD} = 1.42), indicating that most participants felt the system supported effective teamwork. Similarly, participants generally agreed that they \textit{"found it easy to understand my partner's actions"} (\textit{M} = 4.67, \textit{SD} = 1.78), underscoring the clarity and visibility of collaborative interactions.

\subsubsection{Synchronization and Frustration}

Participants also rated their perceived synchronization during the task, yielding more moderate responses. The question \textit{"I felt synchronized with my partner during the task"} had a mean score of 3.92 (\textit{SD} = 1.68), revealing mixed perceptions of how aligned participants felt with their partner’s actions. This variance is particularly notable when compared to satisfaction scores, suggesting that while users could collaborate, there was room for improvement in fostering real-time alignment. In terms of frustration, responses were somewhat polarized. The question \textit{"I feel frustrated when I use the tool and/or the collaborative environment"} received a wider range of responses, with scores spanning from 1 to 6 (\textit{M} = 3.67, \textit{SD} = 1.61). This distribution reflects the diverse user experiences encountered, with a few outliers reporting extreme dissatisfaction, likely stemming from technical issues or unfamiliarity with VR interfaces.

\subsubsection{Visual Quality of Environment and Expectations}

The quality of the final environment was another area where participants' opinions varied. When asked, \textit{"The quality of the environment we created was high,"} the mean score was 3.08 (\textit{SD} = 1.38), indicating that users were not overwhelmingly impressed with the final output. However, the statement \textit{"The tool met my expectations for collaborative creation"} received slightly more favorable responses (\textit{M} = 4.33, \textit{SD} = 1.50), showing that, despite some limitations, the system largely met participants' expectations for a collaborative tool.

\subsubsection{Satisfaction and Future Use}

Regarding satisfaction with outcomes, participants provided mixed feedback. The question \textit{"I am satisfied with the outcome of this tool"} had a mean score of 3.92 (\textit{SD} = 1.31), and the related question \textit{"I was able to achieve what I had in mind with this tool"} had a mean of 2.83 (\textit{SD} = 1.19), suggesting that while participants were generally able to complete their tasks, the final results did not always align with their initial intentions. This could reflect the inherent challenges of prompt-based creation, where scene generation is sometimes constrained by the system's interpretive capabilities. In contrast, more positive feedback was gathered concerning future intent to use the system. The statement \textit{"I would like to use this tool again for similar tasks"} received consistently higher ratings (\textit{M} = 4.75, \textit{SD} = 1.86), with minimal variance, indicating that despite some frustrations, participants saw the potential for future collaborative creation tasks using the system. Despite the lower mean in achieving the final goal, the system was overall well perceived and participants seem to be overall satisfied with the experience. This could hint on further potential of the system and its functionalities that we will discuss later in the paper.

\subsection{System Evaluation Findings}
In addition to the user study, we also conducted a technical evaluation of the Scene Generation using the previously described spatial reasoning dataset. Our primary approach, which integrates visual feedback, was compared against three baselines: one baseline that uses only textual feedback, another following the same set of instructions but lacking visual feedback, and a final baseline that does the same but uses GPT-4 \cite{openai2024chatgpt} instead of GPT-4o. For evaluation, we employed our full pipeline to compare the predicted positions of each asset with the ground truth, which specifies whether an asset should be placed to the left, right, behind, in front, below, or above another asset. To assess accuracy, we programmatically analyzed pairwise interactions between assets by examining the relative coordinates: for x-coordinates, if the predicted position is less than the reference, the asset is correctly placed to the left; for y-coordinates, if the predicted position is less, it is correctly placed in front; and for z-coordinates, if less, the asset is correctly placed below. A placement is considered correct if the predicted relationships between assets align with the ground truth across all pairwise interactions. The accuracy of the method is then calculated as the ratio of correct placements to total relationships.

Using GPT-4 in the pipeline yields an accuracy of 71.4\%. When the underlying model is changed to GPT-4o, we observe a significant increase in accuracy to 83.3\%. We hypothesize that GPT-4o's \textit{"end-to-end"} training across text, vision, and audio modalities, where all inputs and outputs are processed by a single neural network, improves multimodal reasoning, which is especially beneficial in this task \cite{gpt4o}. Simply incorporating feedback with the state of the system in text did not seem to provide much benefit, achieving an accuracy of 83.5\%. Incorporating our vision-based feedback mechanism further enhances the accuracy to 88.8\%. Despite the 3D environment, leveraging explicit 2D visual information allows us to refine the semantic arrangement of assets, improving the system's ability to accurately interpret spatial relationships.

\section{Discussion}

In this section, we dive into the key discussion points that emerged from both developing and evaluating our system. Contributing to emerging research directions in HCI and Human-AI interaction, we propose design implications for similar systems and outline potential directions for future work -- specifically ones relating to AI-determinism, multi-user collaborations in virtual spaces, and fairness in creating such systems.

\subsection{Leaving It Up to Chance: User Intent and AI Interpretation}

The element of randomness in GenAI systems introduces a philosophical dimension to user interaction, particularly in the context of virtual worldbuilding. While some users expressed concerns regarding the misalignment between prompts and outcomes, the concept of "controlled" randomness warrants a deeper conceptual exploration. To what extent should there be room for the unexpected, allowing GenAI to surprise users with its interpretations?

This element of unexpectedness, referred to as \textit{"system's quirks"} by one of our participants, represents occasional deviations from algorithmic determinism. These quirks can manifest as unintentional errors or glitches, yet they also contribute to the richness and diversity of the generated content. The question arises: should users be more explicit in their prompts to minimize these quirks, or should the system be designed to embrace multiple asset selections and interpretations? 

Maintaining the balance between predictability and surprise is crucial for enhancing user engagement and creativity. On one hand, providing users with clear and consistent outcomes can improve their sense of control and satisfaction. On the other hand, allowing for elements of surprise can stimulate creativity and exploration, leading to novel and unexpected results. This duality underscores the importance of designing systems that can accommodate both user preferences and the inherent unpredictability of AI-driven content generation. 

Our findings suggest that users appreciate the element of surprise, as it adds an element of excitement and discovery to the interaction. However, it is essential to ensure that these surprises do not lead to frustration or confusion. Implementing features that allow users to toggle between more deterministic and more random modes of generation could offer a solution, providing flexibility and catering to diverse user needs.

\subsection{Social Collaboration as Virtual Performance}

In collaborative virtual environments, the spontaneous creation of 3D content by multiple users raises several pertinent questions. Who are the creators designing for? What assets, environments, and interactions do they choose to prompt? By assigning an open-ended collaboration task to our study participants, we were able to observe distinct patterns of interaction, creation, and object manipulation within the environment.

Participants often created environments as a form of self-expression, reflecting their visions and poetic choices. At times, they aimed to co-create specific experiences with their partners or reminisce about shared memories from the past. These varied motivations underscore the diverse potential of virtual worlds. 

A multi-user human-AI runtime system like \textit{Social Conjurer} needs to support a wide range of collaborative and creative scenarios without imposing limitations on users. Virtual worlds offer numerous design affordances, and the multiplayer feature is intended to facilitate this expansive potential. The findings from our study highlight the importance of providing users with the freedom to explore and create in an open-ended manner. By fostering an environment that encourages spontaneous and collaborative creation, we can unlock poetic and creative modes of user interaction in virtual spaces.

\subsection{Private vs. Public: Who Owns the Objects?}

In networked virtual environments, the ownership of created assets and scenes is a critical consideration. Networking libraries facilitate the creation of host-client, shared, or private environments. In \textit{Social Conjurer}, a notable limitation arises when objects are spawned: these objects are owned by the client that initiated their creation. Consequently, any client that does not own a specific object is unable to manipulate it, including prompting behaviors or modifying the object from the Unity Editor. The current system's reliance on Photon Fusion 2’s networking framework restricted these assets to client-specific ownership, causing friction when participants attempted to collaboratively modify objects. 

The issue of ownership is fundamentally tied to the core principles of networking. The question of ownership becomes increasingly pertinent as users prompt tools to generate objects within virtual scenes. Should these tools spawn objects for all clients in the network, or should they be restricted to individual users? Furthermore, should modifications made to shared objects be reflected across the entire environment? These questions are inevitable in networked scenes and warrant further investigation in future work on networked 3D spaces augmented with AI-building agents.

\subsection{When Words Fall Short: Linguistic Choices and Multi-Modal Interactions}

In prompt-based systems like \textit{Social Conjurer}, it is crucial to distinguish between different modes of expression and user interaction. Linguistic inputs alone may not suffice to fully capture a user's intentions. Our participants' prompts to make the scene \textit{"look more normal"} or \textit{"a little bit prettier"} reveal a struggle to convey aesthetic refinements through text alone. Incorporating other multi-modal types of input and interactions, such as tools, gestures, and animations, becomes essential. For instance, users might prompt tools to modify existing objects in the scene or directly manipulate objects and the environment through intuitive gestures.

When language falls short, the system should support the expression of abstract intentions and aesthetic visions. This can be achieved through features like layout auto-suggestions and maintaining flexibility in the visual appearances of objects. By doing so, we can help users reach the \textit{"Aha! This is exactly what I wanted!"} moment more efficiently, without the system taking excessive runtime to achieve the desired outcome.

A critical consideration is the trade-off between the time it takes for the system to generate the desired outcome and the time it takes for the user to achieve their intended result. How can we optimize this process? What other types of multi-modal interactions beyond text and tools can facilitate this? And how can we make these interactions collaborative? These are the questions that future work should aim to address.

\subsection{Towards Ethical 3D Worlds}

The ethical considerations of our system encompass several key areas. First, the content and its appropriation. It is essential to implement filters and profanity checks in environments where censorship is necessary, such as applications for under-age users, educational settings, and corporate environments. However, it is equally important to ensure that the tool supports artistic expression, where censorship could limit creativity, such as in creating 3D drawings and graffiti. Additionally, our reliance on third-party 3D asset libraries, such as Objaverse \cite{objaverse}, to search and pull objects into the environment necessitates careful scrutiny to ensure that the content is age-appropriate, diverse, and representative of a variety of objects, people, and experiences. To avoid embedding our own biases into the system, we must consider reducing dependency on these libraries by, for example, using 3D-generating models, such as TripoSR \cite{tochilkin2024triposr},  where appropriate, which also need to be evaluated for AI fairness. Future iterations of systems such as \textit{Social Conjurer} will have to take into account the iterative aspect of creating dynamic interactive scenarios into account in their attempts at filtering, if the system were to be released commercially. This would likely require a video understanding model running in the background to detect potentially unwanted behavior. Questions remain how many camera feeds one might need in arbitrary scene generation cases.

Second, we must ensure that the user interface and experience are equitable. This involves addressing several critical questions: How accessible is the system? Who has access to it? Who owns the objects in the server-client relationship, and why? Ensuring that the virtual power dynamic remains equitable and fair is paramount. The current system assumes that users are able-bodied and neurotypical, capable of easily navigating the interface, the virtual environment, and managing cognitive overload. While there are ways to prompt AI to remove and replace certain textures and assets, we must broaden these applications to ensure that all users can benefit from our system. A potential work is to incorporate customizable accessibility features that allow users to tailor the interface and interactions to their specific needs, thereby enhancing usability of existing tools and frameworks, e.g. \cite{elor2022physical, kurniawan2024virtual}.

Third, when developing computationally expensive systems with generative AI, it is crucial to consider the environmental cost of running these applications. Minimizing the carbon footprint of created virtual worlds, by reducing the number of API calls and optimizing other parameters can significantly lower energy consumption. Implementing efficient algorithms, leveraging renewable energy sources, and optimizing server usage are essential steps in this direction. Additionally, exploring techniques such as model compression and efficient data handling can further contribute to reducing the environmental impact. By prioritizing sustainability in the development process, we can create more eco-friendly virtual environments without compromising on performance or user experience.

\section{Conclusion}

Virtual Reality has sometimes been conceived as an exploration of a potential new methodology for general communications. In this conception, users would become able to make up the content and dynamics of a shared virtual world as it is experienced. Early researchers in VR sometimes referred to this idea as "post-symbolic communication". Natural language is built of symbols which refer to elements of reality, whether physical or logical. Virtual Reality can be directly experienced, so it is possible that a form of communication might be possible in which the expressive flexibility of symbols is retained in a more direct, neo-concrete theater of experience. This is a speculative and difficult idea. If it is ever to be explored, it must first become possible for those experiencing a social virtual world to also create unforeseen types of changes to contents and dynamics during runtime that can be mutually experienced. We have extended the existing body of work showing how GenAI can be applied in runtime to virtual worlds to begin this journey.

In this paper, we presented the \textit{Social Conjurer} system, which lets multiple users collaborate on creating 3D content by prompting virtual worlds at runtime. Using the Unity Game Engine, the users are able to view, modify, and interact with these worlds in a VR headset. In addition to the collaborative world creation, \textit{Social Conjurer} is augmented with multiple GenAI-based modules, which enhance the system's capability to generate interactive environments, tools, and changes to the scene layout using VLMs.

Our research addresses the following questions: How can we develop a real-time, collaborative, spatially-aware system that integrates both LLMs and VLMs to facilitate the co-creation of virtual worlds? How do shared virtual environments influence spontaneous, collaborative world building? What are the challenges and opportunities of using language-based prompts to generate 3D virtual environments?

Through a user study (N=12), we uncovered key insights into how social, multi-user contexts shape the prompting of spatial scenes into existence. Our findings highlight the diverse motivations and interaction patterns among users, ranging from rapid VR prototyping to creating immersive scenarios and reminiscing on shared experiences in virtual worlds. Moreover, discrepancies between user intent and AI interpretation, or controlled randomness, yielded new ideas for prompts in participants' creative processes. Additionally, we discussed design implications and proposed future research directions for similar systems, emphasizing the importance of accessibility, ethical considerations, and optimizing user experience in collaborative virtual environments.

This is a concrete study of what can be done today. Our purpose here is not to speculate about where this type of work might lead in the long term, but this work provides a proof of concept that the modality of post-symbolic communication is becoming explorable by combining advances in GenAI with VR.

\begin{acks}
We would like to thank our study participants and the HCAIX group at Microsoft Research for their continuous help and contributions while running our user study. Additionally, we would like to thank other members of Speaking the World into Existence group, in particular Jennifer Marsman and Sam Earle, for their invaluable feedback during the development of Social Conjurer.
\end{acks}

\bibliographystyle{ACM-Reference-Format}
\bibliography{references}

\newpage

\appendix


 \section{Metaprompts}

 Here we provide the metaprompts and optional structure output requirements for each agent used in Social Conjurer.  A carefully designed metaprompt allows an LLM to follow instructions and learn from in-context demonstrations.
 
 In general, the metaprompts are structured into three sections: introduction, guidelines, and few-shot examples for in-context learning. In some cases, like the Decider, we also provide the JSON schema for the assistant to follow in it's answers.

\subsection{Decider}

We implemented the Decider as an assistant with structured outputs, so that it's answer is restricted to either recommendation of static scene generation or code writing given the user prompt, and we used GPT-4o-mini with the Threads and Assistants API to allow for fast decisions.

The metaptompt for this agent is
\begin{lstlisting}[style=mystyle, caption={Decider metaprompt},label=p1, frame=single]
You are an assistant within a larger framework that takes user prompts and then generates 3D virtual worlds. You are supposed to decide if the user prompt  involves adding only static objects to the scene, such as terrains, environments, or objects, or whether it requires writing code that adds dynamical behavior, animations or other dynamic visual effects. Because the code writing assistant takes longer to complete its job, it is important to only send to it the requests that truly need code writing.
\end{lstlisting}
 \lstset{style=mystyle}

The structured JSON Schema that constrains the outputs is:

\begin{lstlisting}[style=mystyle, caption={Decider JSON Schema},label=p1, frame=single]
{
  "name": "decider",
  "strict": false,
  "schema": {
    "type": "object",
    "properties": {
      "category": {
        "type": [
          "string",
          "null"
        ],
        "description": "Category of prompt: Does it pertain to generating static scenes, or does code need to be written to implement dynamics.",
        "enum": [
          "static_scene",
          "coder"
        ]
      }
    },
    "required": [
      "category"
    ]
  }
}
\end{lstlisting}
 \lstset{style=mystyle}

\subsection{Coder}
The single-user version of the Coder's metaprompt builds upon LLMR Builder in multiple ways, most importantly by providing concrete in-context examples on making VR tools without needing to specify very technical language.
%
%
\begin{lstlisting}[style=mystyle, caption={Coder metaprompt for Conjurer (single-user setting)},label=p1, frame=single]
You are a code-writing assistant that is embedded in the Unity game engine equipped with a runtime C# compiler. I will ask you to generate content in 3d environments, and your task is to write C# code that will implement those requests. You can only respond in code that will compile in C#, and can only add other text in inline comments, like so:
//comment
Do not put the code in a code block, just directly respond with it. That means it should not start with
using UnityEngine;
but rather with just
using UnityEngine;

You are attached to the GameObject "Compiler" in the scene that also has the runtime compiler attached to it. Your generated C# script is automatically attached to this object and immediately executed. Each of your output messages is compiled as a separate script. Make sure to make any elements you might want to reuse public.

# Guidelines you **must** follow
- Each C# script should be a complete and valid class that inherits from the Widget class.
- Each C# script should have a variable called summary that briefly describes the purpose of the script. Write this summary as part of the Start() function.
- Each C# script should also add a Fusion library to the list of used libraries, like so: "using Fusion;".
- Do not assume nor create any prefab references; if you create a public gameObject, make sure to instantiate it within the script, and not rely on me attaching it through the editor.
- Do not apologize when the I point out a mistake and suggest a modification of the code you wrote. Only output the modified code.
- If you want to delete an existing script, use: Destroy(gameObject.GetComponent<script_name>()).
- If it is appropriate to respond to my request by displaying text on the screen, use: Utils.DisplayMessage(message).
- If you're using UI elements in your script, create them yourself as children of the "UI_Canvas" gameObject.
- Avoid using raw numbers in the code. Instead, create a public class variable with the desired value assigned to it.
- ALWAYS make class members public.
- Only write self-contained code. Do not leave anything for me to do.
- When defining more than one classes, they should all be contained within one overarching class.
- When the “Library” is available, put more considerations to the descriptions that guide you on how to write code that uses the relevant skills.
- If I ask you to create a tool, please follow the examples for the tool creation only! Don't try to animate the objects or come up with other behaviors outside of the tool specified.
- DO NOT reference SketchfabLoader under ANY circumstances. Do not use `SketchfabLoader loader = gameObject.GetComponent<SketchfabLoader>()`! Only use the Objaverse API using the instructions described below when loading an object that isn't a primitive.

# VR Interaction Toolkit Instructions
You can create tools that allow VR controllers to interact with objects and terrains in the environment using the XR Interaction Toolkit. Here are the key instructions:
- Making Objects Grabbable: If I request an object to be made grabbable, use the `XRGrabInteractable` component. Ensure the object also has a `Rigidbody` component with appropriate settings (e.g., non-kinematic, gravity enabled).
- Creating Tools: Tools can be created by either spawning new objects or modifying existing objects to turn them into tools. Ensure that each tool or interactable object is fully customizable with public fields to modify behavior or appearance at runtime.

# Utils
You have access to a Utils class that supports various functionalities.
- AddPhysics(gameObject, mass): makes the object physically realistic by attaching colliders and rigidbody to it
- DisplayMessage(message): Use this if it is appropriate to respond to the user's request by displaying text on the screen.

# Objaverse
New models that are not primitives can be added to the scene by using the Objaverse API in combination with Flask server responses. The Objaverse API must be instantiated with `ObjaverseLoader loader = gameObject.GetComponent<ObjaverseLoader>();`. Here’s how to use the API:
- Locate the ObjaverseManager GameObject and get the SceneFlask component. This component is essential for communicating with the Flask server to retrieve the necessary model data:
`GameObject objaverseManager = GameObject.Find("ObjaverseManager");`
- Locate the component on the ObjaverseManager called SceneFlask: `SceneFlask sceneFlask = objaverseManager.GetComponent<SceneFlask>();`
- Specify the scene description in a string called scene and a list of asset names you want to load in a List<string> called asset_names.
- Request JSON Data and Load Models: Use the SceneFlask component to request JSON data from a Flask server and load the models into the scene. This is done by calling a coroutine:
`yield return StartCoroutine(sceneFlask.GetJsonFromFlaskServer(scene, asset_names));`
- Avoid Deprecated Logic. DO NOT reference SketchfabLoader or any deprecated logic. Only use the Objaverse API and the above instructions.

# Prefabs
There are existing prefabs you can use. To reference the prefabs, first initialize the API with `PrefabManager prefab_manager = gameObject.GetComponent<PrefabManager>();`. Below are the available prefabs and their APIs:
## Slider
- `SetName(name)`: sets the name.
- `SetRange(min, max, curr)`: sets the min, max, and current value of the slider.
- `SetPosition(pos)`: sets the position.
- `SetSize(size)`: sets the size.

# Animation
I may ask you to add animations to GameObjects. For this task, you should first find the relevant model and initialize the API with `animation_manager = GetComponent<AnimationManager>();`. It supports the following calls:
- `CreateAnimation(object, clip_name)`: creates an animation that fits the `clip_name` for `object`. Note that you should always call this on the root object you're animating instead of a single joint. For example, if you are animating a bird flapping its wings, do: `CreateAnimation(bird, "flipping_wings")`. Similarly, you should attach animations to the root object. You shouldn't use the animation manager if you're moving simple objects like Unity primitives.

# Examples
## Example
user:
Instructions:
Load a realistic computer desk.
assistant:
using UnityEngine;
using System.Collections;
using System.Collections.Generic;

public class CreateDesk : Widgets
{
    private SceneFlask sceneFlask;

    IEnumerator Start()
    {
        summary = "Loads a computer desk and related assets from the Objaverse using Flask server response";

        // Find the ObjaverseManager GameObject and get the SceneFlask component
        GameObject objaverseManager = GameObject.Find("ObjaverseManager");
        if (objaverseManager == null)
        {
            Utils.DisplayMessage("ObjaverseManager GameObject not found in the scene.");
            yield break;
        }

        sceneFlask = objaverseManager.GetComponent<SceneFlask>();
        if (sceneFlask == null)
        {
            Utils.DisplayMessage("SceneFlask component not found on ObjaverseManager.");
            yield break;
        }

        string scene = "Computer Desk";
        List<string> asset_names = new List<string>
        {
            "desk",
            "chair",
            "computer monitor",
            "keyboard",
            "mouse"
        };

        // Start coroutines to get JSON data from Flask server and load models
        yield return StartCoroutine(sceneFlask.GetJsonFromFlaskServer(scene, asset_names));
    }
}

## Example
user:
Instructions:
Load a couch, a rug, and a siamese cat into the scene.

assistant:
using UnityEngine;
using System.Collections;
using System.Collections.Generic;

public class CreateCouchRugCat : Widgets
{
    private SceneFlask sceneFlask;

    IEnumerator Start()
    {
        summary = "Loads a couch, a rug, and a Siamese cat from the Objaverse using Flask server response, then places the cat on top of the couch";

        // Find the ObjaverseManager GameObject and get the SceneFlask component
        GameObject objaverseManager = GameObject.Find("ObjaverseManager");
        if (objaverseManager == null)
        {
            Utils.DisplayMessage("ObjaverseManager GameObject not found in the scene.");
            yield break;
        }

        sceneFlask = objaverseManager.GetComponent<SceneFlask>();
        if (sceneFlask == null)
        {
            Utils.DisplayMessage("SceneFlask component not found on ObjaverseManager.");
            yield break;
        }

        // Specify the scene and the assets to be loaded
        string scene = "Living Room Scene";
        List<string> asset_names = new List<string>
        {
            "couch",
            "rug",
            "siamese_cat"
        };

        // Start coroutine to get JSON data from Flask server and load models
        yield return StartCoroutine(sceneFlask.GetJsonFromFlaskServer(scene, asset_names));
    }
}

## Example
user:
Instruction: Create a sphere that changes the color of any object it collides with to that of its own.

assistant:

using UnityEngine;

public class ColorChangerSphere : Widgets
{
    public GameObject sphere;

    void Start()
    {
        summary = "This script creates a sphere that can change the colors of objects it collides with to that of its own.";
        // Create a sphere
        sphere = GameObject.CreatePrimitive(PrimitiveType.Sphere);
        sphere.name = "ColoringSphere";
        // make the sphere a color changer
        sphere.AddComponent<ColorChanger>();
    }

    [RequireComponent(typeof(Collider), typeof(Rigidbody), typeof(Renderer))]
    public class ColorChanger : Widgets
    {
        // Start is called

 before the first frame update
        void Start()
        {
            summary = "This script makes the attached object change any object it collides with to have the same color as itself.";
        }

        void OnCollisionEnter(Collision collision)
        {
            Color myColor = GetComponent<Renderer>().material.color;
            // Get the renderer of the other object
            Renderer otherRenderer = collision.gameObject.GetComponent<Renderer>();

            // If the other object has a renderer
            if (otherRenderer != null)
            {
                // Set the color of the other object to the same as this object
                otherRenderer.material.color = myColor;
            }
        }
    }
}

## Example
user:
Instruction: Create a teleporter with two spheres marking its entrance and exit. It should transport object from the entrance to a position slightly below the exit.

assistant:

using UnityEngine;

public class CreateTeleporter : Widgets
{
    public GameObject entrance;
    public GameObject exit;

    void Start()
    {
        summary = "This script creates a teleporter that transports object from the entrance to the exit.";
        // create spheres to mark the entrance and exit of the teleporter
        entrance = GameObject.CreatePrimitive(PrimitiveType.Sphere);
        entrance.name = "TeleporterEntrance";
        exit = GameObject.CreatePrimitive(PrimitiveType.Sphere);
        exit.name = "TeleporterExit";
        exit.transform.position = entrance.transform.position + 10 * entrance.transform.forward;
        // make the sphere entrance
        Teleporter teleporter = entrance.AddComponent<Teleporter>();
        // set the exit
        teleporter.exit = exit.transform;

    }

    [RequireComponent(typeof(Collider), typeof(Rigidbody))]
    public class Teleporter : Widgets
    {
        // The transform of the other end of the teleporter
        public Transform exit;

        // The offset to apply to the teleported object's position
        public Vector3 offset;

        // A flag to prevent infinite loops of teleporting
        private bool isTeleporting;

        void Start()
        {
            summary = "This script teleports collided objects to the exit.";
            offset = new Vector3(0, -1f, 0); //teleport below the exit
            // need isTrigger flag for OnTriggerEnter to work
            gameObject.GetComponent<Collider>().isTrigger = true;
            // the teleporter is not subjected to gravity and other forces, so make it kinematic.
            gameObject.GetComponent<Rigidbody>().isKinematic = true;
        }

        private void OnTriggerEnter(Collider other)
        {
            // Check if the object has the teleport tag and is not already teleporting
            if (!isTeleporting)
            {
                // Set the flag to true
                isTeleporting = true;

                // Get the object's rigidbody if it has one
                Rigidbody rb = other.GetComponent<Rigidbody>();

                // Save the object's velocity and angular velocity
                Vector3 velocity = rb ? rb.velocity : Vector3.zero;
                Vector3 angularVelocity = rb ? rb.angularVelocity : Vector3.zero;

                // Move the object to the exit position plus the offset
                other.transform.position = exit.position + offset;

                // Rotate the object to match the exit rotation
                other.transform.rotation = exit.rotation;

                // Restore the object's velocity and angular velocity
                if (rb)
                {
                    rb.velocity = velocity;
                    rb.angularVelocity = angularVelocity;
                }

                // Set the flag to false
                isTeleporting = false;
            }
        }
    }
}

## Example
user:
Instruction: Create a grabbable object that can be picked up with a VR controller.

assistant:

using UnityEngine;
using UnityEngine.XR;
using UnityEngine.XR.Interaction.Toolkit;

public class GrabbableObject : Widgets
{
    private XRGrabInteractable grabInteractable;
    private Rigidbody rb;

    void Start()
    {
        summary = "This script makes the object grabbable with VR controllers.";
        grabInteractable = gameObject.GetComponent<XRGrabInteractable>() ?? gameObject.AddComponent<XRGrabInteractable>();
        rb = gameObject.GetComponent<Rigidbody>() ?? gameObject.AddComponent<Rigidbody>();
        rb.isKinematic = false;
        rb.useGravity = true;
        ConfigureXRGrabInteractable();
    }

    void ConfigureXRGrabInteractable()
    {
        grabInteractable.interactionLayers = InteractionLayerMask.GetMask("Interactable");
        grabInteractable.throwOnDetach = true;
        grabInteractable.trackPosition = true;
        grabInteractable.trackRotation = true;
    }
}

## Example 
user:
Instruction: Create a tool that can extrude terrain when pointed at it with a VR controller and the trigger is pressed.

assistant:

using UnityEngine;
using UnityEngine.XR;
using UnityEngine.XR.Interaction.Toolkit;

public class TerrainExtruder : MonoBehaviour
{
    private XRGrabInteractable grabInteractable;
    private Rigidbody rb;
    private Terrain terrain;
    public float extrusionHeight = 0.01f;
    public int extrusionRadius = 10;
    public XRNode controllerNode = XRNode.RightHand;
    private GameObject toolObject;
    private bool isToolHeld = false;

    void Start()
    {
        // Summary or description of the script
        Debug.Log("This script makes the object a terrain extruding tool when grabbed.");

        // Find the terrain in the scene
        terrain = FindObjectOfType<Terrain>();
        if (terrain == null)
        {
            Debug.LogError("No Terrain found in the scene!");
            return;
        }

        // Create the tool GameObject
        CreateToolObject();

        // Add necessary components to the tool object
        grabInteractable = toolObject.AddComponent<XRGrabInteractable>();
        rb = toolObject.AddComponent<Rigidbody>();
        rb.isKinematic = false;
        rb.useGravity = true;

        // Configure the XRGrabInteractable settings
        ConfigureXRGrabInteractable();
    }

    void CreateToolObject()
    {
        // Create a new GameObject for the tool
        toolObject = GameObject.CreatePrimitive(PrimitiveType.Cylinder);
        toolObject.name = "TerrainExtruderTool";
        toolObject.transform.localScale = new Vector3(0.1f, 0.5f, 0.1f); // Scale it to resemble a tool

        // Position the tool in front of the player or set a default position
        toolObject.transform.position = Camera.main.transform.position + Camera.main.transform.forward * 0.5f;
    }

    void ConfigureXRGrabInteractable()
    {
        grabInteractable.interactionLayers = InteractionLayerMask.GetMask("Interactable");
        grabInteractable.throwOnDetach = true;
        grabInteractable.trackPosition = true;
        grabInteractable.trackRotation = true;

        grabInteractable.selectEntered.AddListener(OnSelectEntered);
        grabInteractable.selectExited.AddListener(OnSelectExited);
    }

    void OnSelectEntered(SelectEnterEventArgs args)
    {
        // Set the flag to true when the tool is picked up
        isToolHeld = true;
    }

    void OnSelectExited(SelectExitEventArgs args)
    {
        // Set the flag to false when the tool is released
        isToolHeld = false;
    }

    void Update()
    {
        // Check if the tool is held and the trigger is pressed to extrude terrain
        if (isToolHeld && IsTriggerPressed())
        {
            ExtrudeTerrain();
        }
    }

    private bool IsTriggerPressed()
    {
        InputDevice device = InputDevices.GetDeviceAtXRNode(controllerNode);
        return device.TryGetFeatureValue(CommonUsages.triggerButton, out bool isPressed) && isPressed;
    }

    private void ExtrudeTerrain()
    {
        // Logic for extruding the terrain based on the tool's position and extrusion parameters
        Vector3 toolPosition = toolObject.transform.position;
        Vector3 terrainPosition = terrain.transform.position;
        int terrainX = Mathf.FloorToInt((toolPosition.x - terrainPosition.x) / terrain.terrainData.size.x * terrain.terrainData.heightmapResolution);
        int terrainZ = Mathf.FloorToInt((toolPosition.z - terrainPosition.z) / terrain.terrainData.size.z * terrain.terrainData.heightmapResolution);

        float[,] heights = terrain.terrainData.GetHeights(terrainX - extrusionRadius / 2, terrainZ - extrusionRadius / 2, extrusionRadius, extrusionRadius);
        for (int x = 0; x < extrusionRadius; x++)
        {
            for (int z = 0; z < extrusionRadius; z++)
            {
                heights[x, z] += extrusionHeight;
            }
        }
        terrain.terrainData.SetHeights(terrainX - extrusionRadius / 2, terrainZ - extrusionRadius / 2, heights);
    }
}
\end{lstlisting}
 \lstset{style=mystyle}

\subsection{Networked Coder}

The following version of the Coder's metaprompt reflects the use of Unity's networking library Photon Fusion 2 to allow the multi-user capability into the project.

\begin{lstlisting}[style=mystyle, caption={Networked Coder},label=p1, frame=single]
You are a code-writing assistant that is embedded in the Unity game engine equipped with a runtime C# compiler. I will ask you to generate content in 3d environments, and your task is to write C# code that will implement those requests.  You can only respond in code that will compile in C#, and can only add other text in inline comments, like so:
//comment
Do not put the code in a code block, just directly respond with it. That means it should not start with
```csharp
using UnityEngine;
but rather with just
using UnityEngine;

You are attached to the GameObject "Compiler" in the scene that also has the runtime compiler attached to it. Your generated C# script is automatically attached to this object and immediately executed. Each of your output messages is compiled as a separate script. Make sure to make any elements you might want to reuse public.

# Guidelines you **must** follow
- Each C# script should be a complete and valid class that inherits from the NetworkWidget class.
- Each C# script should have a variable called summary that briefly describes the purpose of the script. Write this summary as part of the Start() function.
- Each C# script should also add a Fusion library to the list of used libraries, like so: "using Fusion;".
- Do not assume nor create any prefab references; if you create a public gameObject, make sure to instantiate it within the script, and not rely on me attaching it through the editor.
- Do not apologize when the I point out a mistake and suggest a modification of the code you wrote. Only output the modified code. 
- If you want to delete an existing script, use: Destroy(gameObject.GetComponent<script_name>()). 
- If it is appropriate to respond to my request by displaying text on the screen, use: Utils.DisplayMessage(message).
- If you're using UI elements in your script, create them yourself as children of the "UI_Canvas" gameObject.
- Avoid using raw numbers in the code. Instead, create a public class variable with the desired value assigned to it.
- ALWAYS make class members public.
- Only write self-contained code. Do not leave anything for me to do.
- When defining more than one classes, they should all be contained within one overarching class.
- When the “Library” is available, put more considerations to the descriptions that guide you on how to write code that uses the relevant skills.
- If I ask you to add some kind of behavior to an existing object in the scene, please default to assuming that it's a networked behavior, inhering from the NetworkWidget class, and is using Fusion libraries.
- If I ask you to create a tool, please follow the examples for the tool creation only! Don't try to animate the objects or come up with other behaviors outside of the tool specified.

# Utils
You have access to a Utils class that supports various functionalities.
- AddPhysics(gameObject, mass): makes the object physically realistic by attaching colliders and rigidbody to it
- DisplayMessage(message): Use this if it is appropriate to respond to the user's request by displaying text on the screen.

# Objaverse
New models that are not primitives can be added to the scene by using the Objaverse API, which has to be instantiated with Objaverse loader = gameObject.GetComponent<ObjaverseLoader>();. Here's how you use the API:
- After instantiating the Objaverse API, models can be loaded by calling the method ObjaverseLoader.LoadObjects(model_data_list)
- Each model data in the list has the form "new ModelData("", object_name, reference_object_name, position, scale)". If reference_object_name is not empty, the position of the object is set relative to the reference object.
- You should scale each object according to their real-world sizes. A scale of 1 in Unity corresponds to a size of 1 meter in reality.
- DO NOT make any references to SketchfabLoader, as this logic is deprecated. Please only use the new ObjaverseAPI with the instructions I gave above.

# Examples

## Example 
user:
Instructions:
Load a realistic computer desk.
assistant:
using UnityEngine;
using System.Collections;
using System.Collections.Generic;
using Fusion;

public class CreateDesk : NetworkWidget
{
    IEnumerator Start()
    {
        summary = "loads from objaverse a computer desk";

        // get the script component from the game object
        ObjaverseLoader loader = gameObject.GetComponent<ObjaverseLoader>();

        // create a list of models with labels, positions, and scales
        List<ModelData> models = new List<ModelData>() { new ModelData("", "computer_desk", "", new Vector3(1,0,0), 1f) }; // the desk is located at (1,0,0) and has a scale of 1
        yield return loader.LoadObjects(models);
    }
}

## Example
user:
Scene: There is nothing in the scene.
Library: n/a
Instruction: Create a custom networked drawing tool using mouse input.

assistant:
using System.Collections.Generic;
using System.Collections;
using UnityEngine;
using System;
using Fusion;

public class MouseNetworkDrawTool : NetworkWidget
{
    private LineRenderer currentLine;
    private List<Vector3> points = new List<Vector3>();
    private NetworkObject tool;
    private float lineWidth = 0.05f;
    private float minimumDistance = 0.1f;
    private int batchSize = 5;  // Number of points to batch before sending

    void Start()
    {
        summary = ""This script creates a custom networked drawing tool using mouse."";

        // Tool creation pipeline as a networked gameobject
        try
        {
            Debug.Log(""Sending request to Basic Spawner..."");
            tool = BasicSpawner.OnToolPrompt(); // Always reference this when defining a tool gameObject

            // Start the coroutine to wait for confirmation and then assign behaviors
            StartCoroutine(WaitForObjectAndAssignBehavior(tool));
        }
        catch (Exception e)
        {
            Debug.Log(""Exception occurred: "" + e);
        }
    }

    private IEnumerator WaitForObjectAndAssignBehavior(NetworkObject networkedObject)
    {
        // Ensure networkedObject is non-null
        if (networkedObject == null)
        {
            Debug.LogError(""Networked object is null."");
            yield break;
        }

        // Log initial state
        Debug.Log(""Waiting for networked object to be confirmed..."");

        // Wait until the object is fully confirmed
        while (!networkedObject)
        {
            Debug.Log(""Networked object not confirmed yet..."");
            yield return null; // Wait until the next frame
        }

        // Now the object is confirmed, so you can safely assign behaviors or components
        Debug.Log(""Networked object is confirmed. Assigning behaviors..."");

        AssignBehaviors(networkedObject);
    }

    private void AssignBehaviors(NetworkObject networkedObject)
    {
        Debug.Log(""Behavior assigned!"");

        // These assignments are important to have both clients in sync!
        currentLine = networkedObject.gameObject.AddComponent<LineRenderer>();
        Material lineMaterial = new Material(Shader.Find(""Unlit/Color""));
        lineMaterial.color = Color.red;
        currentLine.material = lineMaterial;
        currentLine.startWidth = lineWidth;
        currentLine.endWidth = lineWidth;
        currentLine.useWorldSpace = true;
        currentLine.positionCount = 0;
        points.Clear();

        if (currentLine == null)
        {
            Debug.LogError(""Start: No LineRenderer found on the object. Please attach a LineRenderer component."");
        }
        else
        {
            Debug.Log(""Start: LineRenderer initialized successfully."");
        }

        // You might want to initialize the points list as well
        points.Clear();
    }

    void Update()
    {
        if (Input.GetMouseButtonDown(0))
        {
            StartDrawing();
        }

        if (Input.GetMouseButton(0) && currentLine != null)
        {
            UpdateDrawing();
        }

        if (Input.GetMouseButtonUp(0))
        {
            EndDrawing();
        }
    }

    void StartDrawing()
    {
        currentLine = tool.gameObject.GetComponent<LineRenderer>();
        Material lineMaterial = new Material(Shader.Find(""Unlit/Color""));
        lineMaterial.color = Color.red;
        currentLine.material = lineMaterial;
        currentLine.startWidth = lineWidth;
        currentLine.endWidth = lineWidth;
        currentLine.useWorldSpace = true;
        currentLine.positionCount = 0;
        points.Clear();
    }

    void UpdateDrawing()
    {
        Vector3 mousePosition = GetMouseWorldPosition();

        if (points.Count == 0 || Vector3.Distance(points[points.Count - 1], mousePosition) >= minimumDistance)
        {
            points.Add(mousePosition);
            currentLine.positionCount = points.Count;
            currentLine.SetPositions(points.ToArray());

            // Send points in batches to reduce the number of RPC calls
            if (points.Count % batchSize == 0)
            {
                RPC_AddPoints(points.GetRange(points.Count - batchSize, batchSize).ToArray());
            }
        }
    }

    void EndDrawing()
    {
        if (points.Count % batchSize != 0)
        {
            // Send any remaining points that weren't sent in the last batch
            int remainingPoints = points.Count % batchSize;
            RPC_AddPoints(points.GetRange(points.Count - remainingPoints, remainingPoints).ToArray());
        }

        RPC_FinalizeLine();
        currentLine = null;
    }

    [Rpc(RpcSources.All, RpcTargets.All)]
    private void RPC_AddPoints(Vector3[] newPoints)
    {
        Debug.Log($""RPC_AddPoints called, adding {newPoints.Length} points."");

        if (currentLine != null)
        {
            points.AddRange(newPoints);
            currentLine.positionCount = points.Count;
            currentLine.SetPositions(points.ToArray());
        }
    }

    [Rpc(RpcSources.All, RpcTargets.All)]
    private void RPC_FinalizeLine()
    {
        Debug.Log($""RPC_FinalizeLine called, finalizing the line."");

        currentLine = null;
    }

    Vector3 GetMouseWorldPosition()
    {
        Vector3 mousePosition = Input.mousePosition;
        mousePosition.z = 10f; // Adjust the Z depth as needed
        return Camera.main.ScreenToWorldPoint(mousePosition);
    }

}

## Example
user: 
Scene: A scene with some arbitrary networked game object.
Library: n/a
Instruction: make the object named "CustomObject" move up and down.

assistant:
using UnityEngine;
using Fusion;

public class NetworkFloatTest : NetworkWidget
{
    private NetworkObject customObject;
    private NetworkTransform networkPosition;
    private float originalY;
    private bool movingUp = true;
    private float moveSpeed = 1.0f;
    private float moveRange = 2.0f;

    void Start()
    {
        summary = "This script makes the CustomObject in the scene float up and down.";
        // Find the networked object called "CustomObject"
        GameObject obj = GameObject.Find("CustomObject");
        if (obj != null)
        {
            customObject = obj.GetComponent<NetworkObject>();
            networkPosition = obj.GetComponent<NetworkTransform>();
            networkPosition.DisableSharedModeInterpolation = true;
            originalY = obj.transform.position.y;
        }
        else
        {
            Debug.LogError("CustomObject not found in the scene.");
        }
    }

    public void Update()
    {
        if (customObject != null)
        {
            // Calculate the new position
            float newY = customObject.transform.position.y;

            if (movingUp)
            {
                newY += moveSpeed * Time.deltaTime;
                if (newY >= originalY + moveRange)
                {
                    newY = originalY + moveRange;
                    movingUp = false;
                }
            }
            else
            {
                newY -= moveSpeed * Time.deltaTime;
                if (newY <= originalY)
                {
                    newY = originalY;
                    movingUp = true;
                }
            }

            Vector3 newPosition = new Vector3(customObject.transform.position.x, newY, customObject.transform.position.z);
            customObject.transform.position = newPosition;
        }
    }
}

## Example
user:
Scene:
Library:
Instruction: Create a primitive cube gameobject. The object is networking through Photon Fusion 2. It should also change color on mouseclick, which should be transferred through an RPC method to other clients. This all should happen within a single script.

assistant:
using UnityEngine;
using Fusion;
using Fusion.Sockets;
using System.Collections.Generic;

public class NetworkedChangeColorTest : NetworkWidget
{
    private Renderer cubeRenderer;

    void Start()
    {
        summary = "This script creates a primitive cube and synchronizes the colors change through an RPC method.";
    }

    void Update()
    {
        // Only allow the host or the owner to process input and change color
        if (Input.GetMouseButtonDown(0))
        {
            Debug.Log("Mouse click.");
            // Raycast from the mouse position
            Ray ray = Camera.main.ScreenPointToRay(Input.mousePosition);
            RaycastHit hit;

            if (Physics.Raycast(ray, out hit))
            {
                // Check if we hit this cube
                if (hit.transform == this.transform)
                {
                    Debug.Log("Hit.");
                    // Change color and notify others
                    Color newColor = new Color(Random.value, Random.value, Random.value);
                    if (Runner.IsRunning && Runner.SessionInfo.IsValid)
                    {
                        RPC_ChangeColor(newColor);
                    }
                }
            }
        }
    }

    [Rpc]
    private void RPC_ChangeColor(Color color)
    {
        Debug.Log("Changing color requested.");
        // Change the color of the cube
        if (cubeRenderer == null)
        {
            cubeRenderer = GetComponent<Renderer>();
        }
        cubeRenderer.material.color = color;
    }
}
\end{lstlisting}
\lstset{style=mystyle}

\subsection{Networked Inspector}

Networked Inspector is responsible for checking the Networked Coder's output accuracy, accommodating for the respective networking libraries.

\begin{lstlisting}[style=mystyle, caption={Networked Inspector},label=p1, frame=single]
You're a meticulous inspector who will evaluate a snippet of Unity C# code. The provided code will be automatically attached to the "Compiler" gameObject, and the Start() function for any classes defined will be executed. Note that "Compiler" contains scripts that are responsible for the behavior of objects in the scene, so it is okay for the code to try and locate the script on the "Compiler" instead of other gameObjects. You need to assess whether the provided code abides by the following set of guidelines.

# Guidelines
- The code assigns each created gameObject a unique name. Note that this doesn't mean the variable names are unique, but rather the gameObject names are unique. GameObjects are named in the following way: GameObject x = new GameObject([Name]), where [Name] is the unique name.
- The code does not add a component in the Start() function of the class with the same name, as that creates an infinite loop. For example, if the class name is Foo, DO NOT use AddComponent<Foo>() in the Start() function.
- The code works as-is without any additional actions from the Unity editor. For example, the code should not declare a public GameObject and expects it to be filled in from the editor screen. However, it is okay for the code to declare a public GameObject but assigns them values later through code. It is only an issue if there is a public variable declared but not assigned any values.
- The following components and functions are not used: NavMesh;
- The script always inherits from NetworkWidget instead of MonoBehaviour. You can assume the Networked_Widgets class exists.
- The script always changes the "summary" variable describing its purpose. Assume it is already declared. It is not useless.
- The script shouldn't contain any code outside of the main class, except statements that start with "using". Ex: "using UnityEngine;" can exist outside of the main class. 
- Make sure any user input gameobjects like hands and controllers have the right names. Always refer to the UserInput string.
- Make sure there are no placeholder functions.
- Unity tags are not used.
- Unity layers are not used.

# Notes
- You can assume "UI_Canvas" always exists in the scene.
- It is always acceptable to declare variables as public, even if they are assigned values in the code.

In addition, the code is allowed to use these methods from the Utils class
# Utils
You have access to a Utils class that supports various functionalities.
- AddPhysics(gameObject, mass): makes the object physically realistic by attaching colliders and rigidbody to it
- PlaceNextTo(base_obj, obj_to_place, offset): place obj_to_place next to base_obj in the direction of offset. The placement will be resolved such that the object's mesh bounding boxes do not overlap.
- DisplayMessage(message): Use this if it is appropriate to respond to the user's request by displaying text on the screen.

# Output format
Your input will be the following:
Scene: ...
Library: ...
Code: ...

Where "Scene" is a brief summary of the existing objects in the Unity scene. "Library" is a relevant skill that is implemented and can be used in the code. "Code" contains the code you need to evaluate.

Your output must be in the format below:

Reasoning: ...
Suggestion: ...
Verdict: ...

Be judicious and write down your step-by-step evaluation of the code under "Reasoning", then propose how this code can be modified so that it meets the guidelines in "Suggestion". Your suggestion should be succinct. Do not include the modified code, just describe how the code should be changed. Finally, "Verdict" should be either "success" or "fail".

# Examples

## Example 
user:
Instructions:
Load a realistic computer desk.
assistant:
using UnityEngine;
using System.Collections;
using System.Collections.Generic;
using Fusion;

public class CreateDesk : NetworkWidget
{
    IEnumerator Start()
    {
        summary = "loads from objaverse a computer desk";

        // get the script component from the game object
        ObjaverseLoader loader = gameObject.GetComponent<ObjaverseLoader>();

        // create a list of models with labels, positions, and scales
        List<ModelData> models = new List<ModelData>() { new ModelData("", "computer_desk", "", new Vector3(1,0,0), 1f) }; // the desk is located at (1,0,0) and has a scale of 1
        yield return loader.LoadObjects(models);
    }
}

## Example
user:
Scene: There is nothing in the scene.
Library: n/a
Instruction: Create a drawing line renderer tool that works with mouse clicks.

assistant:
using System.Collections;
using System.Collections.Generic;
using UnityEngine;
using Fusion;

public class DrawWithMouse : NetworkWidget
{
    private LineRenderer line;
    private Vector3 prevPos;
    private List<Vector3> points = new List<Vector3>();

    [SerializeField]
    private float minDistance = 0.1f;

    // Start is called before the first frame update
    void Start()
    {
        summary = "This script creates a drawing tool with a mouse.";
        line = GetComponent<LineRenderer>();
        line.positionCount = 1;
        prevPos = transform.position;
    }

    // Update is called once per frame
    void Update()
    {
        if (Input.GetMouseButtonDown(0))
        {
            Vector3 currPos = Camera.main.ScreenToWorldPoint(Input.mousePosition);
            currPos.z = 1.16f;

            if (prevPos == transform.position)
            {
                line.SetPosition(0, currPos);
            } else
            {
                line.positionCount++;
                line.SetPosition(line.positionCount - 1, currPos);
            }
            points.Add(currPos); 
            prevPos = currPos;
            
            if (Runner.IsRunning && Runner.SessionInfo.IsValid)
            {
                RPC_ReceivePoint(currPos);
            }

        }
    }

    [Rpc]
    void RPC_ReceivePoint(Vector3 point)
    {
        Debug.Log("Received point: " + point);
        points.Add(point);
        UpdateLineRenderer();
    }

    void UpdateLineRenderer()
    {
        line.positionCount = points.Count;
        line.SetPositions(points.ToArray());
    }
}

## Example
user: 
Scene: A scene with some arbitrary networked game object.
Library: n/a
Instruction: make the object move up and down.

assistant:
using Fusion;
using UnityEngine;

public class NetworkedMoveUpDown : NetworkWidget
{
    public float amplitude = 6.0f; // The height of the movement
    public float frequency = 3.0f; // The speed of the movement
    public NetworkTransform networkPosition;

    private Vector3 startPosition;

    void Start()
    {
        summary = "This scripts creates a networked move up and down behavior.";
        // Store the starting position of the GameObject
        startPosition = transform.position;
        networkPosition = GetComponent<NetworkTransform>();
    }

    public override void FixedUpdateNetwork()
    {
        float newY = startPosition.y + Mathf.Sin(Time.time * frequency) * amplitude;
        transform.position = new Vector3(startPosition.x, newY, startPosition.z);
        networkPosition.Teleport(new Vector3(startPosition.x, newY, startPosition.z));
    }
}

## Example
user:
Scene:
Library:
Instruction: Create a primitive cube gameobject. The object is networking through Photon Fusion 2. It should also change color on mouseclick, which should be transferred through an RPC method to other clients. This all should happen within a single script.

assistant:
using UnityEngine;
using Fusion;
using Fusion.Sockets;
using System.Collections.Generic;

public class NetworkedChangeColorTest : NetworkWidget
{
    private Renderer cubeRenderer;

    void Start()
    {
        summary = "This script creates a primitive cube and synchronizes the colors change through an RPC method.";
    }

    void Update()
    {
        // Only allow the host or the owner to process input and change color
        if (Input.GetMouseButtonDown(0))
        {
            Debug.Log("Mouse click.");
            // Raycast from the mouse position
            Ray ray = Camera.main.ScreenPointToRay(Input.mousePosition);
            RaycastHit hit;

            if (Physics.Raycast(ray, out hit))
            {
                // Check if we hit this cube
                if (hit.transform == this.transform)
                {
                    Debug.Log("Hit.");
                    // Change color and notify others
                    Color newColor = new Color(Random.value, Random.value, Random.value);
                    if (Runner.IsRunning && Runner.SessionInfo.IsValid)
                    {
                        RPC_ChangeColor(newColor);
                    }
                }
            }
        }
    }

    [Rpc]
    private void RPC_ChangeColor(Color color)
    {
        Debug.Log("Changing color requested.");
        // Change the color of the cube
        if (cubeRenderer == null)
        {
            cubeRenderer = GetComponent<Renderer>();
        }
        cubeRenderer.material.color = color;
    }
}

## Example 
user:
Instructions:
Load a realistic computer desk and a red chair using objaverse.
assistant:
using UnityEngine;
using System.Collections;
using System.Collections.Generic;
using Fusion;

public class CreatDeskChairCat : NetworkWidget
{
    IEnumerator Start()
    {
        summary = "loads from objaverse a computer desk and a red chair";

        // get the script component from the game object
        ObjaverseLoader loader = gameObject.GetComponent<ObjaverseLoader>();

        // create a list of models with labels, positions, and scales
        List<ModelData> models = new List<ModelData>() { new ModelData("", "computer_desk", "", new Vector3(1,0,0), 1f) }; // the desk is located at (1,0,0) and has a scale of 1
        yield return loader.LoadObjects(models);
    }
}
\end{lstlisting}
\lstset{style=mystyle}

\subsection{Network Widget Class}

NetworkWidget is a variation of the Widget class, written specifically for the prompted objects that are being spawned over the network with Photon Fusion 2. Below is the implementation of the class.

\begin{lstlisting}[style=mystyle, caption={Network Widget Class},label=p1, frame=single]
using System.Collections;
using System.Collections.Generic;
using UnityEngine;
using Fusion;

public class NetworkWidget : NetworkBehaviour
{
    //public string name;
    public string summary;
    // Start is called before the first frame update
    void Start()
    {
        summary = "This networked widget is a test.\n";
        updateWidgetDB();
    }

    // Update is called once per frame
    void Update()
    {

    }

    public void updateWidgetDB()
    {
        Debug.Log(summary);
    }
}
\end{lstlisting}
\lstset{style=mystyle}

\subsection{Environment Module}
Here we collect the metaprompts used for the various agents that together make up the Environment Generation submodule. These agents use the Assistants and Threads OpenAI API and were mostly implemented using GPT-4o and GPT4 models in our experiments.
\subsubsection{Low Poly Terrain Metaprompt}

The metaprompt for the low poly environment generation that selects from some simple presets is:

\begin{lstlisting}[style=mystyle, caption={Low Poly Terrain Metaprompt},label=p1, frame=single]
Given the following environments: 'default,' 'dune,' 'farmland,' 'canyon,' 'mountain,' and 'forest,' choose the one that best matches the user's input. Provide only the environment name in your response.

#Examples

## Example 1
User Input: create a cow
Response: farmland

## Example 2
User input: create an office
Response: default
\end{lstlisting}
 \lstset{style=mystyle}

\subsubsection{Realistic Terrain Metaprompt}

The metaprompt for the realistic environment generation that selects from a real-world location is:

\begin{lstlisting}[style=mystyle, caption={Realistic Metaprompt},label=p1, frame=single]
The user has provided a description for a virtual scene they want to create in a VR environment, including specific objects they want to place in the scene. Based on the description, identify a real-world location on Earth that would provide an interesting and suitable terrain for this virtual scene, with a focus on user interaction and object placement. The location should include a flat or gently sloped area suitable for a user to stand on and for placing objects, while avoiding areas directly on top of buildings or other structures. Ensure that the terrain is not too steep or too rugged. If the terrain might contain water, ensure there is an area with higher land suitable for people to stand on and lower land where water can naturally be placed. Prioritize open land areas where structures could be realistically placed.

Instructions:
1. Analyze the description to determine the type of terrain or environment the user wants.
2. Consider geographical features such as mountains, forests, deserts, or coastal areas that match the description.
3. Identify a location on Earth with these features.
4. Ensure the location includes a flat or gently sloped area that is suitable for user interaction and object placement.
5. Avoid selecting areas that are directly on top of existing buildings or other large structures; prioritize open land.
6. Ensure the terrain is not too steep or too rugged, but instead has gentle slopes or flat areas.
7. Return the latitude and longitude of this location in the format {latitude, longitude}.

Example Output: {46.5, 10.5}

Note: Please provide only the coordinates in the specified format without any additional text.
\end{lstlisting}
 \lstset{style=mystyle}

\subsubsection{Material Metaprompt}

The metaprompt for low poly terrain materials that selects from a list of material assets is:

\begin{lstlisting}[style=mystyle, caption={Material Metaprompt},label=p1, frame=single]
You are tasked with selecting the most appropriate Unity material for a virtual environment based on a user's description. You have the following materials available:
1. dark_grass_terrain
2. dirt_terrain
3. extraterrestrial_terrain
4. grass_terrain
5. sand_terrain
6. snow_terrain

Instructions:
- Analyze the user's input, which describes an object or scene they would like created in a virtual environment.
- Choose the material that best matches the description from the list above.
- If none of the materials are relevant, respond with default.
- Ensure your response is the exact name of the material (e.g., sand_terrain) or default.

User Input: '{User's input here}'

Example:
If the user input is 'Create a scene with a sandy beach with palm trees,' respond with sand_terrain.
If the user input is 'Design an office scene,' respond with default.

Output the material name only, without additional explanation.
\end{lstlisting}
 \lstset{style=mystyle}

\subsubsection{Terrain Layer Metaprompt}

The metaprompt for realistic terrain layers that selects from a list of terrain layer assets is:

\begin{lstlisting}[style=mystyle, caption={Terrain Layer Metaprompt},label=p1, frame=single]
You are tasked with selecting the most appropriate Unity material for a virtual environment based on a user's description. You have the following materials available:
1. Black_Sand_TerrainLayer
2. Grass_A_TerrainLayer
3. Grass_B_TerrainLayer
4. Grass_Dry_TerrainLayer
5. Grass_Moss_TerrainLayer
6. Grass_Soil_TerrainLayer
7. Heather_TerrainLayer
8. Muddy_TerrainLayer
9. Pebbles_A_TerrainLayer
10. Pebbles_B_TerrainLayer
11. Pebbles_C_TerrainLayer
12. Rock_TerrainLayer
13. Sand_TerrainLayer
14. Snow_TerrainLayer
15. Soil_Rocks_TerrainLayer
16. Tidal_Pools_TerrainLayer

Instructions:
- Analyze the user's input, which describes an object or scene they would like created in a virtual environment.
- Choose the material that best matches the description from the list above.
- If none of the materials are relevant, respond with 'default'.
- Ensure your response is the exact name of the material (e.g., Sand_TerrainLayer) or 'default'.

User Input: '{User's input here}'

Example:
- If the user input is 'Create a scene with a sandy beach with palm trees,' respond with Sand_TerrainLayer.
- If the user input is 'Design an office scene,' respond with 'default'.

Output the material name only, without additional explanation.
\end{lstlisting}
 \lstset{style=mystyle}

\subsubsection{Water Metaprompt}

The metaprompt for deciding whether to include water in the scene:
\begin{lstlisting}[style=mystyle, caption={Terrain Layer Metaprompt},label=p1, frame=single]
You are tasked with determining whether it makes sense for a scene to include water based on a user's description.

Instructions:
- Analyze the user's input, which describes an object or scene they would like created in a virtual environment.
- If the description suggests the presence of water (e.g., beaches, lakes, rivers, rain, etc.), respond with 'true'.
- If the description does not suggest the presence of water, respond with 'false'.

User Input: '{User's input here}'

Example:
- If the user input is 'Create a scene with a beach,' respond with 'true'.
- If the user input is 'Design a mountain landscape,' respond with 'false'.

Output 'true' or 'false' only, without additional explanation.
\end{lstlisting}
 \lstset{style=mystyle}

\subsubsection{Skybox Metaprompt}

The metaprompt for skyboxes that selects from a list of skybox assets is:
\lstinputlisting[style=mystyle, caption={Skybox Metaprompt},label=p1, frame=single]{appendix/metaprompts/skybox_metaprompt.txt}
 \lstset{style=mystyle}

\subsection{Spatial Reasoning module}
Here we collect the metaprompts for the various steps involved in obtaining assets and placing them into the scene. We provide the Python code used to generate the metaprompts dynamically by injecting strings into the prompt at a specified location.

\subsubsection{Asset Retrieval}
The script below returns a pre-selected number of objects given a description of a scene. It is straightforward to generalize this to the case where the number is not specified too.
\begin{lstlisting}[style=mystyle, caption={Python script generating the asset retrieval metaprompt},label=p1, frame=single]
Return {num} assets that would be present in the given scene
**Examples**
8 assets in an underwater world with coral reef:
Coral reef
Clownfish1
Clownfish2
Shipwreck
Jellyfish
Sea Turtle
Seaweed
12 assets in a space station intererior scene:
Control Panel
Laptop
Astronaut
Astronaut2
Airlock Door
Floating Screwdriver
Floating Wrench
Bunk
Medical Kit
Oxygen tanks
Treadmill
**Guidelines** 
Make sure each asset is seperated by only a new line (don't number) and only refers to one thing
{scene}
\end{lstlisting}
 \lstset{style=mystyle}

 \subsubsection{Layout Arrangement}
 As a first pass, we make a call to GPT-4o for a proposal of arrangement of objects in the scene, again in the form of a python script called by the Flask app:

 \lstinputlisting[style=mystyle, caption={Python script generating the first pass layout metaprompt},label=p1, frame=single]{appendix/metaprompts/layout_prompt.txt}
 \lstset{style=mystyle}

\subsubsection{Layout Update} 
Once the initial layout is proposed and we have the updated bounding boxes from downloaded assets, we update the positions of objects in the scene by calling the below metaprompt, plotting the results and feeding them back for several passes (3 in out experiments) to GPT-4o:

 \lstinputlisting[style=mystyle, caption={Python script generating the updated layout metaprompt},label=p1, frame=single]{appendix/metaprompts/update_metaprompt.txt}
 \lstset{style=mystyle}

We then make calls to the VLM to obtain the desired rotations of objects in the scene, by rendering them in Trimesh and then passing the metaprompt below to GPT-4o:

 \lstinputlisting[style=mystyle, caption={Python script generating the metaprompt for obtaining the orientation of each object in the scene},label=p1, frame=single]{appendix/metaprompts/orientation_prompt.txt}
 \lstset{style=mystyle}

\newpage
\section{Spatial Reasoning Evaluation}
The methodology we follow to create the spatial reasoning dataset is detailed in Section 5.4. We expand on this by providing specific prompts and examples from the dataset. We first generate textual descriptions of scenes with various spatial relations:
\begin{lstlisting}[style=mystyle, caption={Scene creation prompt},label=p1, frame=single]
Generate {num} scene descriptions, each containing at least {relations} relational descriptions (e.g., "the person is to the left of the building" or "the street is above the subway"). The relational descriptions must use only the following directions:
- Below, Above (Z-axis)
- Left, Right (X-axis)
- Front, Behind (Y-axis)
\end{lstlisting}
We then use OpenAI's Structured Outputs API to extract a scene graph representation from a given scene description. We use the following as the system content for the call:
\lstinputlisting[style=mystyle, caption={Scene Graph Extraction metaprompt},label=p1, frame=single]{appendix/metaprompts/scene_graph_extraction.txt}
 \lstset{style=mystyle}
With the output class structure defined by:
 \lstinputlisting[style=mystyle, caption={Scene graph class structure },label=p1, frame=single]{appendix/metaprompts/scene_graph_structure.txt}
 \lstset{style=mystyle}
This results in examples such as:
\lstinputlisting[style=mystyle, caption={Dataset Examples},label=p1, frame=single]{appendix/metaprompts/scene_dataset_examples.txt}
The final distribution of spatial relations is provided in Table \ref{tab:spatial-relationships}.

\begin{table}[ht]
\centering
\begin{tabular}{|c|c|}
\hline
\textbf{Spatial Relationship} & \textbf{Count} \\ \hline
Above  & 164  \\ \hline
Behind & 111  \\ \hline
Below  & 161  \\ \hline
Front  & 111  \\ \hline
Left   & 146  \\ \hline
Right  & 147  \\ \hline
\end{tabular}
\caption{Distribution of Spatial Relationships}
\label{tab:spatial-relationships}
\end{table}

\newpage
\section{User Study}

\subsection{Participant Demographics}
\begin{table}[h!]
\begin{tabular}{|r|l|l|l|l|l|}
\hline
\rowcolor[HTML]{000000} 
\multicolumn{1}{|l|}{\cellcolor[HTML]{000000}{\color[HTML]{FFFFFF} ID}} & {\color[HTML]{FFFFFF} Professional Occupation} & {\color[HTML]{FFFFFF} Age} & {\color[HTML]{FFFFFF} Gender} & {\color[HTML]{FFFFFF} GenAI Experience} & {\color[HTML]{FFFFFF} AR/MR/VR Experience} \\ \hline
1                                                                       & PhD Student                                    & 25-34                      & Male                          & Very unfamiliar                         & Slightly unfamiliar                        \\ \hline
2                                                                       & N/A                                            & 18-24                      & Male                          & Familiar                                & Unfamiliar                                 \\ \hline
3                                                                       & Research Intern                                & 18-24                      & Male                          & Familiar                                & Slightly familiar                          \\ \hline
4                                                                       & Computer Science PhD                           & 25-34                      & Male                          & Familiar                                & Familiar                                   \\ \hline
5                                                                       & PhD Student                                    & 25-34                      & Male                          & Very familiar                           & Slightly familiar                          \\ \hline
6                                                                       & Intern                                         & 18-24                      & Female                        & Slightly familiar                       & Neutral                                    \\ \hline
7                                                                       & Software Engineer                              & 18-24                      & Female                        & Slightly familiar                       & Familiar                                   \\ \hline
8                                                                       & Graduate Student                               & 25-34                      & Male                          & Very familiar                           & Slightly familiar                          \\ \hline
9                                                                       & Hardware and Systems                           & 25-34                      & Female                        & Slightly familiar                       & Very unfamiliar                            \\ \hline
10                                                                      & Research Intern                                & 18-24                      & Female                        & Familiar                                & Slightly familiar                          \\ \hline
11                                                                      & Graduate Student                               & 25-34                      & Female                        & Very Familiar                           & Unfamiliar                                 \\ \hline
12                                                                      & PhD Student                                    & 25-34                      & Female                        & Slightly familiar                       & Slightly familiar                          \\ \hline
\end{tabular}
\caption{Demographics of user study participants.}
\label{tab:participant-demographics}
\end{table}

\subsection{Survey Questions}
 Rate these statements (All 1 to 7 Likert scale from Strongly disagree to Strongly agree):
 
 \begin{itemize}
     \item The tool facilitated effective collaboration with my partner.
     \item I found it easy to understand my partner's actions.
     \item I felt synchronized with my partner during the task.
     \item I feel frustrated when I use the tool and/or the collaborative environment. 
     \item I am satisfied with the outcome of this tool.
     \item I was able to achieve what I had in mind with this tool.
     \item I am satisfied with the final environment we created.
     \item The tool met my expectations for collaborative creation.
     \item The quality of the environment we created was high.
     \item I would like to use this tool again for similar tasks.
 \end{itemize}

Short response questions:
\begin{itemize}
    \item In what scenarios would you find it valuable to use a tool like the one you interacted with during this session?
    \item Are there any situations where you would prefer not to use a tool like the one you interacted with during this session? If so, why?
    \item What did you like and dislike about the tool you interacted with today?
    \item Brainstorm positive and negative implications of your vision. Describe them here.
\end{itemize}

\subsection{Verbal Questions}
\begin{itemize}
    \item Can you describe the vision depicted in your concept sketch?
    \item How did collaborating on one device compare to collaborating across two devices? Which did you enjoy more?
\end{itemize}

\end{document}